\documentclass{aa}  

\usepackage{graphicx}
\usepackage{txfonts}
\usepackage[T1]{fontenc}
\usepackage{amsmath}	
\usepackage[flushleft]{threeparttable} 
\usepackage{subcaption}
\usepackage{caption}
\usepackage{comment}
\usepackage[colorlinks=true,linkcolor=blue,citecolor=blue,urlcolor=blue]{hyperref}%
\usepackage{url}
\usepackage{mathrsfs}

\usepackage[switch]{lineno}

\newcommand{\msun}{\mbox{$M_{\odot}$}}

\newcommand{\nick}{\mbox{$^{56}$Ni}}
\newcommand{\cob}{\mbox{$^{56}$Co}}
\newcommand{\fe}{\mbox{$^{56}$Fe}}

\newcommand{\FeI}{Fe~{\sc i}}
\newcommand{\FeII}{Fe~{\sc ii}}
\newcommand{\FeIII}{Fe~{\sc iii}}

\newcommand{\CoII}{Co~{\sc ii}}
\newcommand{\CoIII}{Co~{\sc iii}}

\newcommand{\SiII}{Si~{\sc ii}}

\setcitestyle{citesep={,}}

\begin{document}

   \title{ZTF SN Ia DR2: The secondary maximum in Type Ia supernovae}

   \subtitle{}
   \author{M. Deckers $^{1}$\fnmsep\thanks{E-mail: deckersm@tcd.ie}    
           \and K. Maguire$^{1}$
           \and L. Shingles$^2$
           \and G. Dimitriadis$^1$
           \and M. Rigault$^3$
           \and M. Smith$^{4}$
           \and A. Goobar$^5$
           \and J. Nordin$^{6}$
           \and J. Johansson$^5$
           \and M. Amenouche$^7$
           \and U. Burgaz$^{1}$
           \and S. Dhawan$^8$
           \and M. Ginolin$^3$
           \and L. Harvey$^1$
           \and W. D. Kenworthy$^5$
           \and Y. -L. Kim$^4$
           \and R. R. Laher$^9$
           \and N. Luo$^{1}$
           \and S. R. Kulkarni$^{10}$
           \and F. J. Masci$^{9}$
           \and T. E. Müller-Bravo$^{11, 12}$
           \and P.~E.~Nugent$^{13, 14}$
           \and N. Pletskova$^{15}$
           \and J. Purdum$^{16}$
           \and B. Racine$^{17}$
           \and J. Sollerman$^{18}$
           \and J. H. Terwel$^1$
          }

   \institute{School of Physics, Trinity College Dublin, College Green, Dublin 2, Ireland
   \and GSI Helmholtzzentrum f\"{u}r Schwerionenforschung, Planckstraße 1, 64291 Darmstadt, Germany
   \and Univ Lyon, Univ Claude Bernard Lyon 1, CNRS, IP2I Lyon/IN2P3, UMR 5822, F-69622, Villeurbanne, France
   \and Department of Physics, Lancaster University, Lancaster LA1 4YB, UK
   \and  Oskar Klein Centre, Department of Physics, Stockholm University, SE-10691 Stockholm, Sweden
   \and Institut für Physik, Humboldt-Universität zu Berlin, Newtonstr. 15, 12489 Berlin, Germany
   \and National Research Council of Canada, Herzberg Astronomy \& Astrophysics Research Centre, 5071 West Saanich Road, Victoria, BC V9E 2E7, Canada
   \and  Institute of Astronomy and Kavli Institute for Cosmology, University of Cambridge, Madingley Road, Cambridge CB3 0HA, UK
   \and IPAC, California Institute of Technology, 1200 E. California Blvd, Pasadena, CA 91125, USA
   \and 247-17 Caltech, Pasadena, CA 91125, USA
   \and Institute of Space Sciences (ICE, CSIC), Campus UAB, Carrer de Can Magrans, s/n, E-08193 Barcelona, Spain
   \and Institut d’Estudis Espacials de Catalunya (IEEC), E-08034 Barcelona, Spain
   \and Lawrence Berkeley National Laboratory, 1 Cyclotron Road MS 50B-4206, Berkeley, CA, 94720, USA
   \and Department of Astronomy, University of California, Berkeley, 501 Campbell Hall, Berkeley, CA 94720, USA
   \and Department of Physics, Drexel University, Disque Hall, Office No. 808 32 S. 32nd St., Philadelphia, PA 19104
   \and Caltech Optical Observatories, California Institute of Technology, Pasadena, CA 91125, USA
   \and Aix Marseille Université, CNRS/IN2P3, CPPM, Marseille, France
   \and Oskar Klein Centre, Department of Astronomy, Stockholm University, SE-10691 Stockholm, Sweden
        }

   \date{Received 15 April 2024; accepted 26 June, 2024}

 
  \abstract
  {Type Ia supernova (SN~Ia) light curves have a secondary maximum that exists in the $r$, $i$, and near-infrared filters. The secondary maximum is relatively weak in the $r$ band, but holds the advantage that it is accessible, even at high redshift. We used Gaussian Process fitting to parameterise the light curves of 893 SNe Ia from the Zwicky Transient Facility's (ZTF) second data release (DR2), and we were able to extract information about the timing and strength of the secondary maximum. We found $>5\sigma$ correlations between the light curve decline rate ($\Delta m_{15}(g)$) and the timing and strength of the secondary maximum in the $r$ band. Whilst the timing of the secondary maximum in the $i$ band also correlates with $\Delta m_{15}(g)$, the strength of the secondary maximum in the $i$ band shows significant scatter as a function of $\Delta m_{15}(g)$. We found that the transparency timescales of 97 per cent of our sample are consistent with double detonation models, and that SNe Ia with small transparency timescales ($<$~32 d) reside predominantly in locally red environments. We measured the total ejected mass for the normal SNe~Ia in our sample using two methods, and both were consistent with medians of $1.3\ \pm \ 0.3$ and $1.2\ \pm\ 0.2$ \msun. We find that the strength of the secondary maximum is a better standardisation parameter than the SALT light curve stretch ($x_1$). Finally, we identified a spectral feature in the $r$ band as \FeII, which strengthens during the onset of the secondary maximum. The same feature begins to strengthen at $<$~3 d post maximum light in 91bg-like SNe. Finally, the correlation between $x_1$ and the strength of the secondary maximum was best fit with a broken line, with a split at $x_1^0\ =\ -0.5\ \pm\ 0.2$, suggestive of the existence of two populations of SNe~Ia.}

  \titlerunning{ZTF SN Ia DR2: The secondary maximum}
    \authorrunning{M. Deckers}
   \maketitle
%

\section{Introduction}

Although there is general consensus that Type Ia supernovae (SNe~Ia) originate from the thermonuclear explosions of white dwarfs (WDs) in binary systems, there is still disagreement about the explosion mechanisms and progenitor scenarios \citep[see ][for comprehensive reviews]{Hillebrandt2013, Maoz2014, Ruiter2020, Jha2019, Liu2023a}. The light curves of SNe~Ia around peak light have been extensively studied because in order to use SNe~Ia as cosmological distance indicators \citep{Riess1998, Perlmutter1999, Scolnic2018}, they are standardised using empirical relations between light curve shape, brightness, and colour around maximum light in the optical \citep{Pskovskii1977, Phillips1993, Hamuy1996, Tripp1998}. Around 2--3 weeks post maximum light, SN~Ia light curves begin to rise again to a secondary bump in the optical $i$ band and at near-infrared (NIR) wavelengths \citep{Elias1981}. A shoulder is seen in the optical $r$ band at similar phases. This phenomenon is called the secondary maximum, and is significantly less well studied than the primary maximum due to the fainter magnitudes at these phases combined with the difficulties associated with observing in the NIR. Although less well understood, the secondary maximum encapsulates important information about the physical parameters of the explosion. \cite{Dhawan2016} has shown that the secondary maximum can be used to estimate the amount of \nick\ produced in the explosion, and \cite{Papadogiannakis2019a} has shown the properties of the secondary maximum to be linked to the transparency timescale and total ejected mass in the explosion. Moreover, it has been shown that the secondary maximum in the $i$ and NIR bands can be used to standardise SN~Ia light curves in a similar manner as the primary maximum \citep{Nobili2005, Shariff2016}, although this does not significantly reduce the Hubble residual scatter \citep{Shariff2016}.

The secondary maximum is thought to be caused by the ionisation transition of iron-group elements (IGEs) in the ejecta from doubly- to singly-ionised once the ejecta temperature drops below $\sim$7000K. The increased NIR emissivity of singly, compared to doubly ionised IGEs causes the NIR flux to increase \citep{Pinto2000c, Kasen2006, Kasen2007, Dhawan2015, Hoeflich2017}. This explanation was corroborated by the identification of a strengthening \FeII\ spectral feature in the $i$ band of SN 2014J \citep{Jack2015}, coinciding with the onset of the secondary maximum. Slower evolving light curves with higher peak absolute magnitudes tend to have hotter ejecta and consequently, later secondary maxima \citep{Kasen2006, Blondin2015}. Fainter and faster evolving SNe~Ia have earlier secondary maxima \citep{Hamuy1996, Kasen2006, Taubenberger2017}. The existence of the secondary maximum relies on at least some level of abundance stratification, meaning that explosions with large scale mixing are not expected to show a secondary maximum \citep{Kasen2006}. Moreover, asymmetric ejecta could lead to a viewing angle dependence in the properties of the secondary maximum \citep{Kasen2006}. The progenitor metallicity also affects the secondary maximum through its impact on the amount of stable iron produced, with increased progenitor metallicity leading to an earlier secondary maximum \citep{Kasen2006}.

Some of the more peculiar sub-types of SNe~Ia lack a secondary maximum entirely \citep{Turatto1996, Hoeflich2002, Gonzalez-Gaitan2014, Dhawan2017b, Ashall2020, Lu2021, Hoogendam2022, Dimitriadis2023}. In the case of 91bg-like SNe~Ia or SNe~Iax, the lack of secondary maximum is explained by a very low ejecta temperature, which causes the secondary maximum to merge with the primary peak \citep{Kasen2006, Blondin2015, Taubenberger2017, Galbany2019}. A similar effect has been seen in the late-time light curve of 91bg-like SN 2021qvv \citep{Graur2023}, which lacked a NIR plateau. The NIR plateau occurs between 70--500 d for normal SNe~Ia and is the result of a shift in the dominant ionisation state of IGEs, resembling the secondary maximum \citep{Deckers2023}. On the other extreme end of the SN~Ia absolute luminosity scale, the very bright 03fg-like SNe~Ia are thought to lack a secondary maximum because the secondary re-brightening is obscured by the interaction with a dense circumstellar material (CSM) or envelope \citep{Taubenberger2011, Taubenberger2017, Dimitriadis2021, Dimitriadis2023}. The sub-class of 02es-like SNe~Ia similarly lack a secondary maximum, and \cite{Hoogendam2023} suggest that this sub-class originates from the same progenitor scenario as 03fg-like SNe~Ia, and that they similarly lack a secondary maximum due to interaction with a dense envelope.


\cite{Papadogiannakis2019a} study the secondary maximum of 422 SNe~Ia provided by Carnegie Supernova Project \citep[CSP-I,][]{Contreras2010}, CfA supernova programme \citep{Hicken2009}, the Palomar Transient Factory (PTF), and the intermediate Palomar Transient Factory \citep[iPTF,][]{Rau2009}. Out of these surveys, only PTF/iPTF are untargeted, which makes up about half their sample. They find correlations between the properties of the secondary maximum in the $r$ band and $s_{BV}$ \citep[a proxy for light curve stretch measured by SNooPy,][]{Burns2014} as well as the transparency timescale ($t_0$), which can be directly compared to predictions from explosion models, highlighting the power of the secondary maximum. 

In this paper, we build upon the analysis by \cite{Papadogiannakis2019a} by performing a large scale study of the secondary maximum in the $r$ and $i$ bands for 893 SNe~Ia provided by the Zwicky Transient Facility's (ZTF) second data release \citep[DR2;][]{Rigault2024}. With our large spectroscopic data set, we aim to confirm which specific spectral features are producing the secondary maximum in the $r$ band. We also investigate if the correlations found by \cite{Papadogiannakis2019a} between the secondary maximum and other light curve parameters hold for our larger sample. Finally, with 893 SNe~Ia, DR2 contains many other sub-types beyond just normal SNe~Ia, we investigate the secondary maximum properties of the more peculiar SNe~Ia to test whether their properties can be connected to the normal SN~Ia population through the secondary maximum as was shown by \cite{Li2022} for the transitional SN 2012ij.

In Sect. \ref{data} we introduce the DR2 data set. We describe how we fit and analyse our light curves, introduce average spectral templates, and summarise the final samples analysed in this paper in Sect. \ref{methods}. The results of our light-curve fits are presented in Sect. \ref{results}, and we provide a detailed discussion in Sect. \ref{discussion}. Lastly, we summarise our results in Sect. \ref{conclusions}.

\section{Data}\label{data}

In Section \ref{sample} we describe the ZTF DR2 dataset. In Section \ref{lc_data}, we describe the coverage cuts we applied to ensure we only fit well sampled light curves. 

\subsection{The sample}\label{sample}

For this investigation, we use data from the ZTF survey \citep[ZTF,][]{Bellm2019, Graham2019, Masci2019, Dekany2020}. ZTF is a large field-of-view, galaxy-untargeted transient survey that has discovered thousands of transients since first light in 2018. In particular, we will be investigating the second SN~Ia data release \citep[DR2;][]{Rigault2024}, which contains SN~Ia data of 3628 spectroscopically confirmed events from the first three years of operations (spanning 2018--2020). An overview of the sample statistics and technical details are described in \cite{Smith2024}.

We place a redshift cut at $z \leq 0.06$ because the sample is considered spectroscopically complete for normal SNe~Ia up to $z \leq$0.06 \citep{Amenouche2024}. We note that the fainter sub-types in the sample will not all be discovered to this redshift \citep[see][for further discussion]{Dimitriadis2024}. Placing a redshift cut at $z\leq $0.06 reduces the initial sample size from 3628 to 1584 SNe Ia. 

Additional cuts on the sample are placed in Sects. \ref{sample}, \ref{lc_data}, \ref{gp_fits_method}, \ref{second_max_method}, and \ref{second_strength_method}. We summarise the final samples in Sect. \ref{finalsamples} and in Table \ref{cuts_summary}. 

\begin{table}
\centering
\caption{A summary of cuts applied to produce our final sample.}\label{cuts_summary}
\begin{threeparttable}
\begin{tabular}{ l|l|l } 
\hline \\[-0.8em]
Criterion & No. of SNe & Removed\\
\hline \\[-0.8em]
DR2 & 3628 & -\\
$z \leq 0.06$ & 1584 & 2044\\
Light curve coverage conditions$^a$& 976 & 608\\
GP length scale $<$ 44 d &  973 & 3 \\
$\sigma_{t_2(r)}<5.9$ d & 928 &  45\\
$\sigma_{\mathcal{F}_{r_2}}<0.05$ & 898 &  30\\
Manual inspection of outliers & 893 & 5 \\
Cosmological sub-sample$^b$ & 783 & 110\\
\hline
\end{tabular}
\begin{tablenotes}
    \footnotesize 
    \item [a] Light curve coverage conditions require detections in at least two filters, in at least one filter pre-max and at least one filter post-max. At least two detections pre-peak and at least two detections post peak. A minimum of seven detections across all filters, and a minimum of four detections between +10 and +40 d in the $r$ band.
    \item [b] Sub-sample of SNe~Ia that pass the following cuts: classified as normal or 91T-like, $|x_1|<3$, $|\sigma_{x_1}|\leq 1$, $-0.2\leq c \leq 0.8$, $|\sigma_{c}|\leq 0.1$, $|\sigma_{t_0}|\leq 1$, and fit-probability $>\ 1e^{-7}$ (see Sect. \ref{finalsamples}).
\end{tablenotes}
\end{threeparttable}
\end{table}

\subsection{Light curve data}
\label{lc_data}

In this study we use light curves which were obtained by the ZTF survey in three optical filters (\textit{gri}). We will perform Gaussian Process (GP) fits to characterise these light curves, which work best when the photometry is well sampled. Therefore, we apply quality cuts to the light curve data. We require each light curve to have detections in at least two filters, and in at least one of the same filter pre- and one post-maximum light. We also require a minimum of two detections pre-peak and two post-peak in any filter, and in total a minimum of seven detections across all filters. These criteria are applied to the stacked light curves, meaning that two data points on one epoch only count as a single detection. These light curve quality criteria ensure that we are able to fit the light curves around maximum light to estimate $t_{0_x}$ (the time of maximum in a specific filter, \textit{x}). We also require a minimum of four detections between +10 and +40 d in the $r$ band. This criterion is adopted from \cite{Papadogiannakis2019a} and ensures we are able to adequately constrain the secondary maximum. If there are less than four detections between +10 and +40 d in the $i$ band, we keep the light curve in the sample but we do not attempt to measure the parameters of the secondary maximum in the $i$ band. The aforementioned light curve coverage cuts reduce the size of the sample from 1584 to 976 SNe~Ia.

\section{Analysis}\label{methods}

In Sect. \ref{gp_fits_method}, we describe the GP fits to our light curves and explain our motivation for the choice of parameters \citep[see e.g.,][for details on GP processes]{Rasmussen2006}. In Sections \ref{second_max_method} and \ref{second_strength_method}, we describe the parametrisation of the timing and strength of the secondary maximum, respectively, while in Sect. \ref{uncertainties_k_corrections}, we discuss the impact of {K-corrections} on the parameters we measure. We describe why a number of SNe~Ia were cut from the sample after we inspected outliers by eye, and then summarise all the cuts we applied to arrive at the final two samples used in this analysis. Finally, we introduce a number of spectral templates which we use to constrain the spectral evolution during the secondary maximum.

\subsection{Fitting light curves using Gaussian Processes}\label{gp_fits_method}

To characterise the evolution of the \textit{r}-band light curves, we perform GP fits to the data. GPs are a data-driven, non-parametric method of estimating an underlying function behind data. The same method was implemented by \cite{Papadogiannakis2019, Papadogiannakis2019a} and \cite{Pessi2021} to characterise the \textit{r}- and $i$-band light curves of SN~Ia, respectively. GPs were also used for \textsc{piscola} \citep{Muller-Bravo2022}, a light curve fitter, and \textsc{avocado} \citep{Boone2019}, a photometric classifier.

\begin{figure}
    \centering
    \includegraphics[width=9cm]{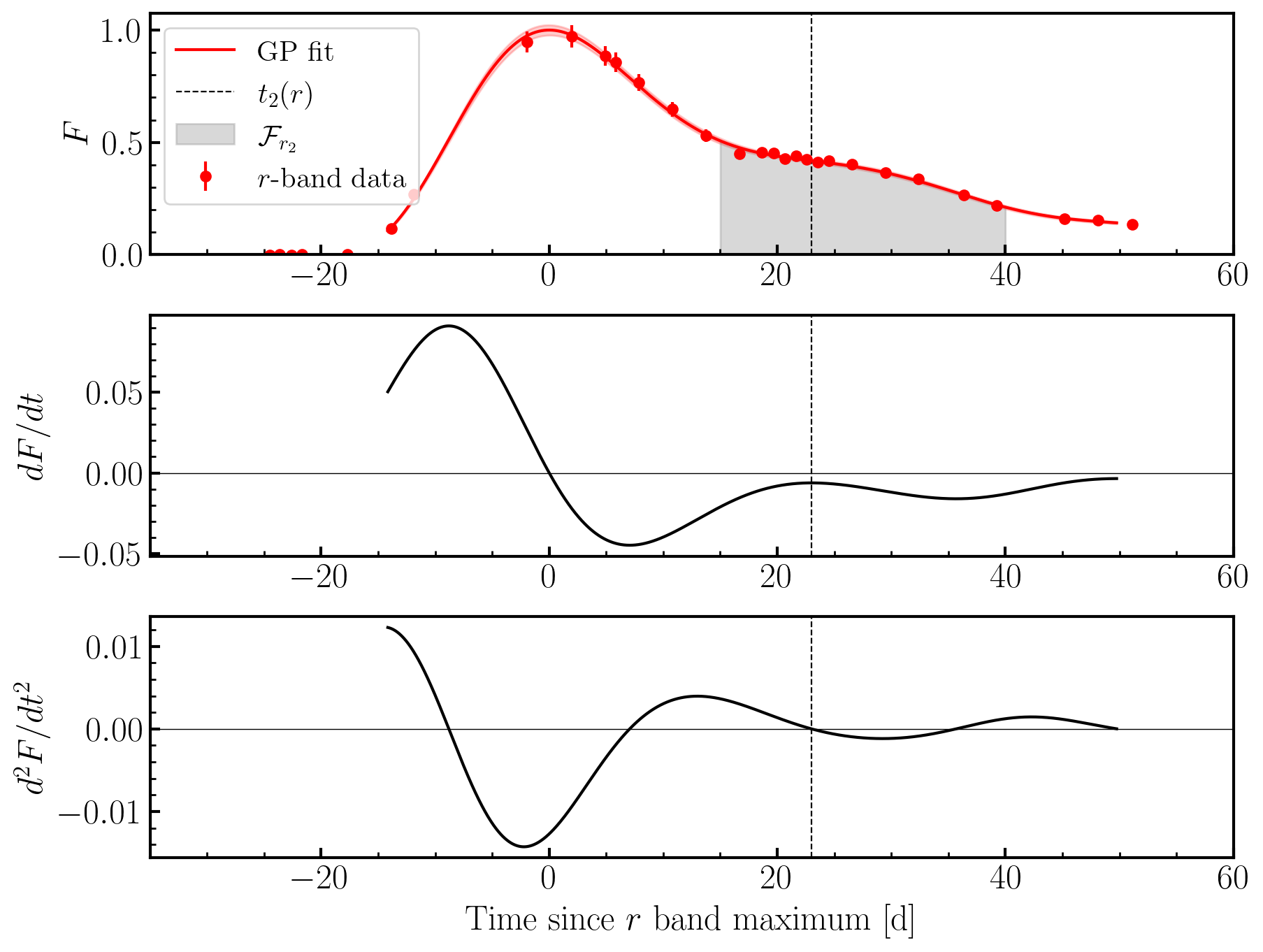}
    \caption{\textbf{Top:} Example of GP fit to a $r$-band light curve. The GP fit is shown as a red line and the $1\sigma$ uncertainty is shown as the red shaded region. The area over which the integrated flux under the secondary maximum, $\mathcal{F}_{r_2}$, is calculated is indicated by the grey shaded region, and $t_2(r)$ by the black dashed line. \textbf{Middle:} The first derivative of the GP light curve. This light curve has a secondary shoulder rather than a secondary maximum because the first derivative is not equal to zero. \textbf{Bottom:} The second derivative of the GP light curve, showing that the time of the onset of the secondary maximum is the time at which the second derivative is equal to zero, also known as an inflection point. }
    \label{GP_fit}
\end{figure}

We use the python package \textsc{scikit-learn} to implement our GP fits, which we perform in 2D in order to fit data across different filters simultaneously. We fit the light curves in 2D to maximise the amount of data available, which helps to constrain the timing of the primary maximum \citep{Dimitriadis2024}. Our analysis focuses primarily on the $r$ band but we also analyse the $i$ band, which generally has less available data than the $r$ band. By fitting the different filters simultaneously we increase the quality of the $i$ band light-curve fits. We fit our nightly stacked ZTF light curves in flux space so that we are able to include non-detections, which could not be used if we were fitting in magnitude space. An uncertainty floor of 2.5, 3.5, and 6 per cent were applied to the flux values in the $g$, $r$, and $i$ band, respectively \citep{Smith2024}. We fit a phase range of $-$15 -- 50 d with respect to maximum light. The fluxes are normalised to peak, as is standard for GP fits to ease maximum likelihood estimation \citep{Rasmussen2006}. An example of an average GP fit of a SN Ia in our sample is shown in the top panel of Fig. \ref{GP_fit}.

\begin{figure*}
    \centering
    \begin{subfigure}{0.49\textwidth}
    \centering
    \includegraphics[width=\textwidth]{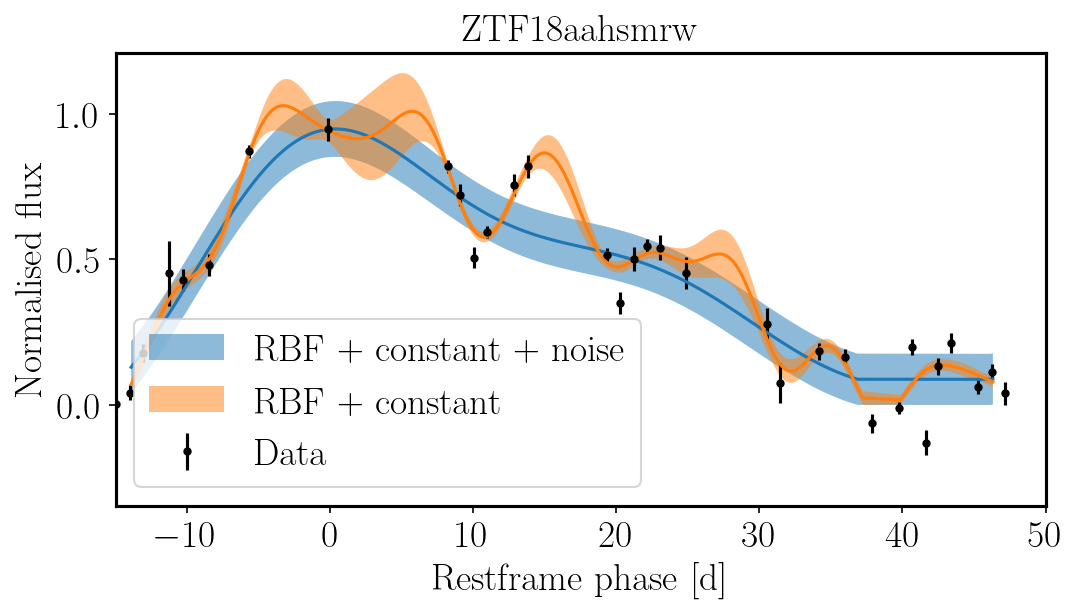}
    \end{subfigure}
    \hfill
    \begin{subfigure}{0.49\textwidth}
    \centering
    \includegraphics[width=\textwidth]{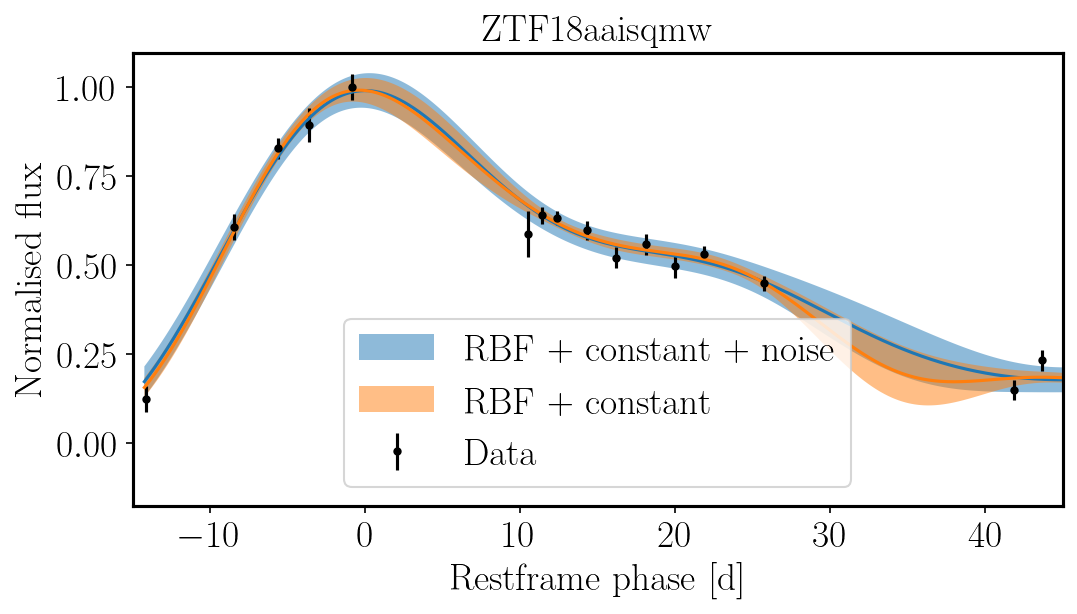}
    \end{subfigure}
    \hfill
    \begin{subfigure}{0.49\textwidth}
    \centering
    \includegraphics[width=\textwidth]{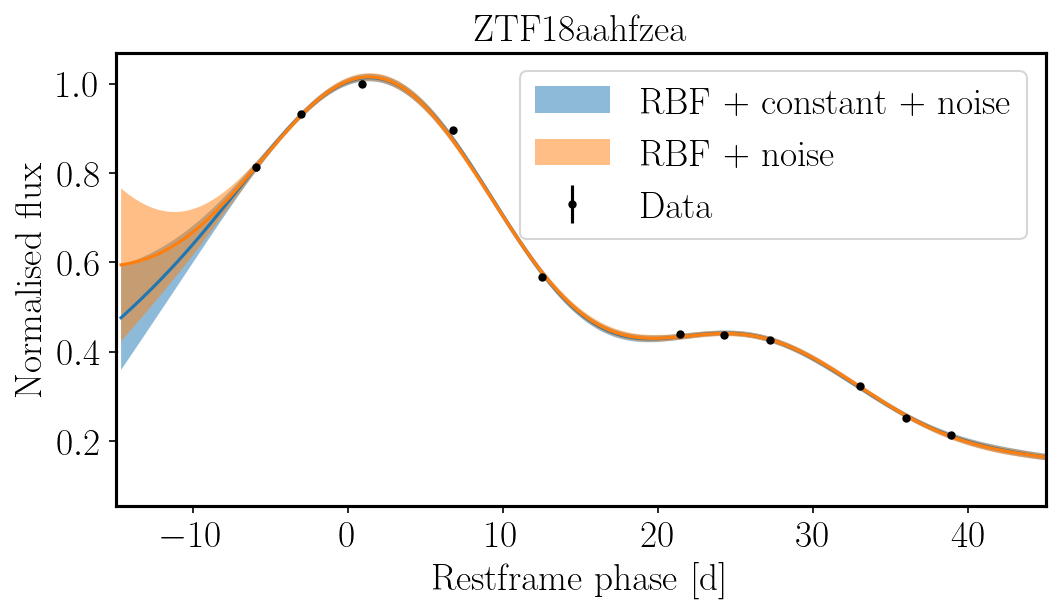}
    \end{subfigure}
    \hfill
    \begin{subfigure}{0.48\textwidth}
    \centering
    \includegraphics[width=\textwidth]{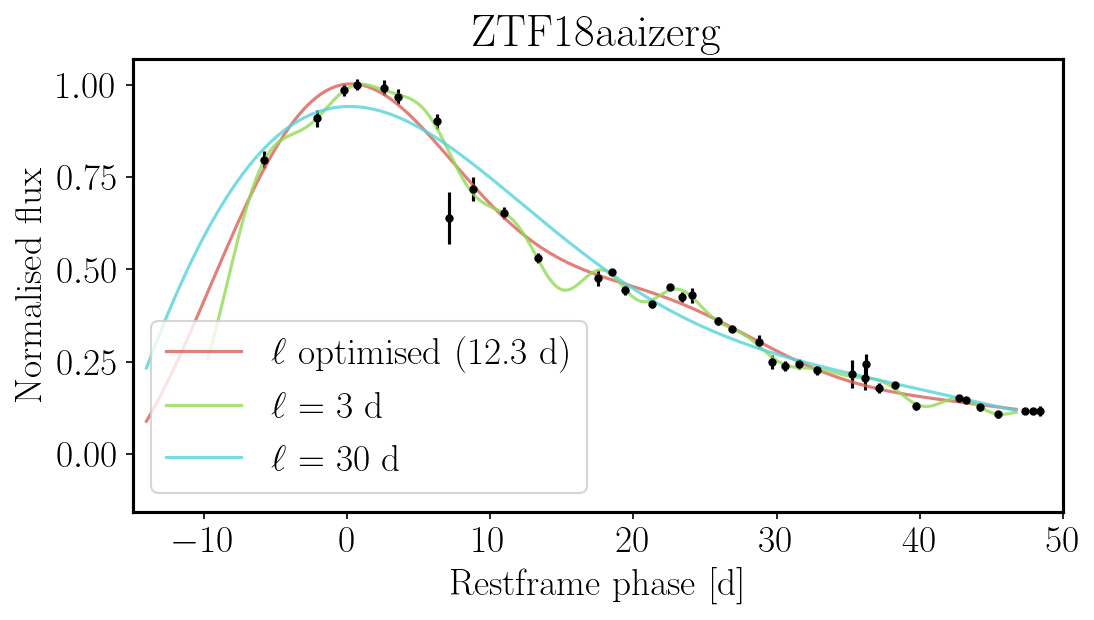}
    \end{subfigure}
        \caption{Impact of various kernel and length scale choices on a GP fit of a ZTF SN Ia light curve. \textbf{Top left:} If the uncertainties in the data are underestimated, not including a white noise kernel will result in unrealistic wiggles in an attempt to fit all the points. \textbf{Top right:} Even if the uncertainties are not underestimated in the data, a lack of a white noise kernel can result in an underestimation of the uncertainty of the GP fit, as is visible between 10--30 d in this light curve. \textbf{Bottom left:} With limited data at the edge of the fitting regions, not including a constant kernel can result in deviating behaviour at the edge regions. \textbf{Bottom right:} We optimise the length scale during the fit, which in this case resulted in $\ell$ = 12.3 d. By setting a length scale, the likelihood is higher to either over-fit if the length scale is too short ($\ell$ = 3 d), or to under-fit and miss important features if the length scale is too long ($\ell$ = 30 d). The ideal value for the length scale will depend on the cadence of the data, which varies across our sample.}\label{gp_kernel_tests}
\end{figure*}

The choice of kernels is important for GP fitting. We test both the radial basis function (RBF) kernel and the Matern kernel. The RBF kernel is an infinitely differentiable, stationary kernel, meaning that it produces smooth output functions \citep{Rasmussen2006, Duvenaud2014}, and depends only on the distance of two data-points, which is parameterised by a length scale parameter ($\ell$), and not on their absolute values. The Matern kernel is also stationary, but it has an additional parameter $\nu$, which controls the smoothness of the function. The Matern kernel only produces smooth, infinitely differentiable functions at particular values of $\nu$ \citep{Rasmussen2006, Duvenaud2014}. We find that the Matern kernel overfits the ZTF light curves, resulting in spurious wiggles in the model. We therefore implement the RBF kernel for our GP fits. \cite{Pessi2019, Pessi2021} also opted for a RBF kernel, whereas \cite{Papadogiannakis2019a} opted for a Matern kernel. 

We test how adding additional kernels (white noise, constant) to our GP model impacts the fits (see Fig. \ref{gp_kernel_tests}). In the top left panel of Fig. \ref{gp_kernel_tests} we highlight that if a light curve has large gaps in the data, a GP without a white noise kernel will over-fit the data resulting in unphysical fluctuations in the flux predictions. The top right panel of Fig. \ref{gp_kernel_tests} shows that a GP without a white noise kernel is likely to under-predict the uncertainties, as is shown between 10--30 d. The bottom left panel of Fig. \ref{gp_kernel_tests} shows that excluding the constant kernel can result in deviating behaviour at the edges of the function, particularly if there is limited data in these regions. As a result of these tests, we opted to include a constant and a white noise kernel. 

We also test the impact of the length scale of the time axis of the RBF kernel on the GP fits. In the bottom right panel of Fig.~\ref{gp_kernel_tests} we show the fit for a typical event in terms of data coverage, by fixing the length scale to either a small value ($\ell$ = 3 d) or a large value ($\ell$ = 30 d). The small value results in over-fitting of the data, while the large value results in an overly smoothed function that does not capture the crucial variation of SN~Ia light curves around the secondary maximum. We also show a fit where the length scale is optimised without bounds in the bottom right panel of Fig.~\ref{gp_kernel_tests}. For this example event the length scale after optimisation is 12.3 d, and the fit captures the evolution of the light well without over-fitting. 

Motivated by our tests, we choose to use the RBF kernel, combined with a constant and a noise kernel. We optimise the length scale of the RBF kernel for each SN, since the optimum values will depend on the cadence of each individual light curve. We find a range of length scales between 4--70 d with a median of 10 d. Upon inspection by eye, light curves with very long length scales are not well fit, so we remove any SNe~Ia from our sample that are fit with a length scale $>44$ d (rejecting measurements which are further than 3$\sigma$ away from the median of the distribution). This cut removes three out of 973 SNe~Ia from our sample. We implement non-informative priors on the hyper parameters of the other kernels, and run an optimisation for each SN fit. 

We estimate the uncertainties on the timing and strength of the secondary maximum (Sect. \ref{second_max_method} and \ref{second_strength_method}) by resampling. We perform 1000 iterations where we perturb the stacked flux data points within their Gaussian uncertainties and rerun the GP fits. We repeat our measurements and take the standard deviation across the 1000 iterations as the uncertainty. 

\begin{figure}
    \centering
    \includegraphics[width=8.5cm]{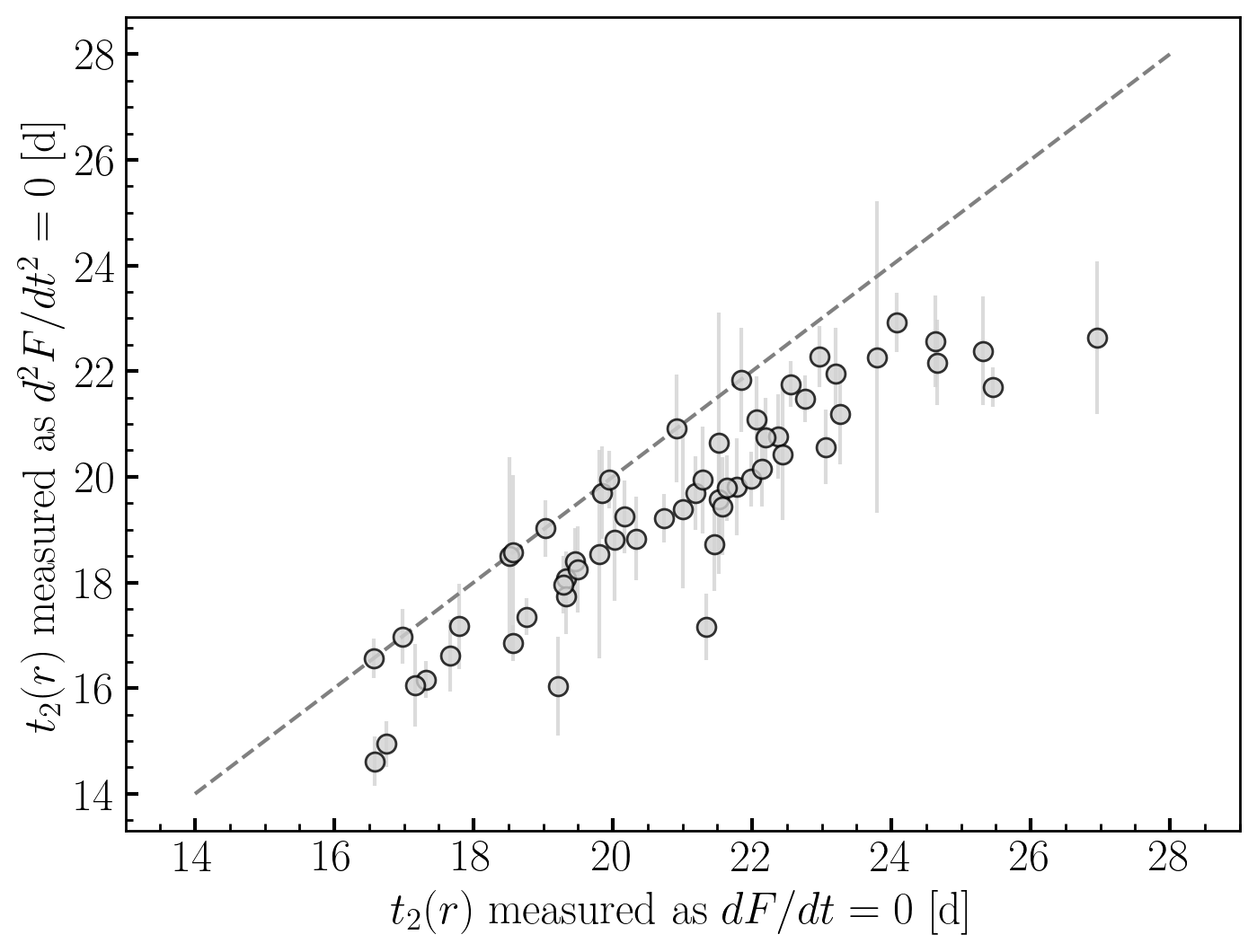}
    \caption{The distribution of $t_2(r)$ for the 63 SNe~Ia with a detected true bump (i.e. $dF/dt\ =\ 0$ exists), measured either as the local maximum ($dF/dt\ =\ 0$, x-axis) or the inflection point ($d^2F/dt^2\ =\ 0$, y-axis). We systematically find a lower $t_2(r)$ if measured as the point of inflection  vs. the local maximum in the light curve.}
    \label{bump_vs_shoulder}
\end{figure}

\begin{figure*}
    \centering
    \includegraphics[width=15.5cm]{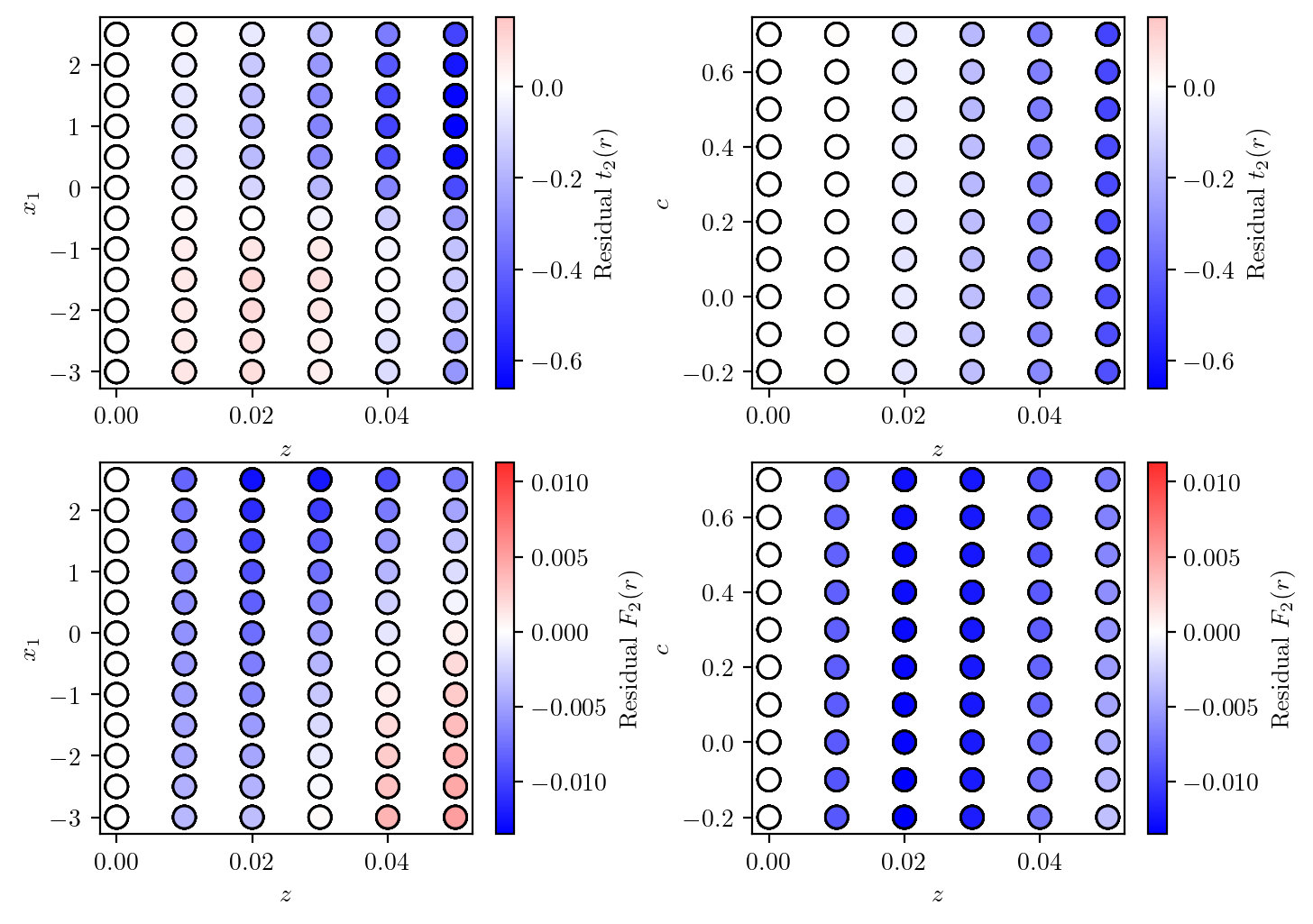}
    \caption{Impact of K-corrections on $t_2(r)$ (top panel) and $\mathcal{F}_{r_2}$ (bottom panel), as a function of redshift and $x_1$ (left) and $c$ (right). The colour of the data points represents the residual between the measured parameter in the SALT3 model at that point and the same model at $z$ = 0. Overall the impact of K-corrections is small, with a maximum residual on $\mathcal{F}_{r_2}$ of 0.014 and on $t_2(r)$ of 0.66 d. }
    \label{kcor}
\end{figure*}

\subsection{Determining the time of the secondary maximum}\label{second_max_method}

To determine the time of the secondary maximum, we trace the GP light curves with a univariate spline and calculate the first and second derivatives. The secondary maximum is expected to occur in the range of 10--40~d with respect to maximum light \citep{Papadogiannakis2019a}. \cite{Papadogiannakis2019a} search this range for where the first derivative is equal to zero and the second derivative is negative, which would correspond to a local maximum in the light curve. Although this would work in the $i$ band, the secondary maximum is less prominent in the \textit{r} band and it is usually better characterised as a shoulder rather than a local maximum. In order to constrain a shoulder in the \textit{r}-band light curve, we locate the first inflection point (where the second derivative is equal to zero, see Fig. \ref{GP_fit}) in the range 13--35 d, which we define as $t_2(r)$. We reduced the upper bound of the range from 40 d to 35 d because \cite{Papadogiannakis2019a} found no secondary maxima later than 35 d, and we find that there is often limited data at this range which can result in an unexpected behaviour of the GP function. We reject any measurements of $t_2(r)$ with an uncertainty more than 3$\sigma$ away from the median of the uncertainty distribution ($\sigma_{t_2}\ >\ 5.9$ d), to eliminate objects with poor GP fits, which removes 45 SNe~Ia. 

We find 20 91bg-, seven 02cx-, and three 02es-like SNe Ia (including SN 2019yvq), with $t_2(r)$ >20 d. These sub-types are not expected to show strong, late secondary maxima, making them stand out significantly from the rest of the sample. We have checked all their light curves by eye and find that their light curves flatten as they approach zero flux around 20 d. Because the absolute flux is very low at these epochs, small fluctuations are picked up in the derivatives and can be classified as inflection points, but these are not a real secondary shoulders/maxima. This is not a problem for most of the normal SNe Ia in the sample because they decline less rapidly, and therefore do not have such low flux values in the window where we search for the secondary maximum. Because the light curves of these objects are well fit by the GP function, we do not remove these SNe Ia from our sample, but instead we only remove their measured $t_2(r)$ values. The only exceptions are SN 2018jpk (an 02cx-like SN Ia) and SN 2019ner (an 02es-like SN Ia) which do appear to have a real secondary maximum at 20.9 and 20.6 d, respectively. We have kept both of their $t_2(r)$ values in the sample.

Our $t_2(r)$ values are better described as the onset of the shoulder, rather than the peak of the secondary maximum (see Fig. \ref{GP_fit}). \cite{Papadogiannakis2019a} also measured the inflection point, but only for light curves where they could not locate a true secondary maximum ($dF/dt$ = 0). We opt to use the point where $d^2F/dt^2\ =\ 0$ (inflection point) for all light curves to ensure we are consistent across the sample because we are interested in relative $t_2(r)$ values. We are able to find a secondary maximum for 808 SNe~Ia, and 60 of these could be characterised as having a true bump ($dF/dt$ = 0)\footnote{In this paper we refer to the secondary shoulder in the $r$ band, $t_2(r)$, as a secondary maximum, even if $dF/dt$ = 0 does not exist in the light curve.}. A direct comparison of our $t_2(r)$ values to those presented in \cite{Papadogiannakis2019a} is hindered by the fact that they do not specify which $t_2(r)$ values correspond to $dF/dt$ = 0 vs. $d^2F/dt^2\ =\ 0$.

In Fig. \ref{bump_vs_shoulder} we highlight the importance of using a consistent definition of $t_2(r)$. We measure $t_2(r)$ only for SNe~Ia with an identified bump ($dF/dt\ =\ 0 $), using both methods. We find that the median difference between the inflection point $d^2F/dt^2\ =\ 0$ and the local maximum in the light curve ($dF/dt\ =\ 0 $) is 1.9 d, with the inflection point generally occurring earlier. 

We also measure $t_2$ using the inflection point ($d^2F/dt^2\ =\ 0$) for the $i$ band ($t_2(i)$). We find that 103 of the $i$ band light curves could be characterised as having a true bump ($dF/dt\ =\ 0$). This is more than in the $r$ band, which is not surprising because the secondary maximum is known to be stronger in the $i$ band. The median offset between the point at which $dF/dt\ =\ 0$ and $d^2F/dt^2\ =\ 0$ is 4.6 d, which is larger than in the $r$ band. This is because the secondary maximum is more prominent and lasts longer in the $i$ band \citep[see fig. 10,][]{Kasen2006}, and as a result the time between the onset and the peak of the secondary maximum is greater.

\subsection{Constraining the strength of the secondary maximum} \label{second_strength_method}

We use the integrated flux under the secondary maximum, $\mathcal{F}_{r_2}$, to quantify the strength of the secondary maximum (see shaded region in the top panel of Fig.~\ref{GP_fit}). The same metric was used in \cite{Krisciunas2001} and \cite{Papadogiannakis2019a} to quantify the strength of the secondary maximum in the \textit{i} and \textit{r} bands, respectively. $\mathcal{F}_{r_2}$ is calculated by normalising the flux to peak in the $r$ band, integrating the flux between +15 and +40 d (where rest-frame phase is defined relative to peak in the $r$ band) and dividing by the length of the interval (25 d). We reject any measurements of $\mathcal{F}_{r_2}$ with an uncertainty more than 3$\sigma$ away from the median of the uncertainty distribution ($\sigma_{\mathcal{F}_{r_2}}\ >\ 0.05$) in order to eliminate objects with poor GP fits, which removes 30 SNe~Ia.

\subsection{Uncertainties due to K-corrections}\label{uncertainties_k_corrections}

At high redshifts, it is important to apply K-corrections because the observed $r$-band will begin to resemble rest-frame $g$-band, where no secondary maximum is expected. Unfortunately, this is not straightforward due to a lack of reliable templates at the phases we are investigating (+15 to +40~d post peak). Moreover, some of the more peculiar sub-types present in our sample do not have sufficient template coverage to calculate reliable K-corrections at any phases.

We estimate the impact of K-corrections using the SALT3 \citep[Spectral Adaptive Light-curve Template;][]{Kenworthy2021} template in the \textsc{sncosmo} library \citep{Barbary2022}. We produce a grid of light curves which resemble the normal SNe Ia in our sample, with $-3\ \leq\ x_1\ \leq \ 3$ in steps of 0.5 and $-0.2\ \leq\ c\ \leq \ 0.8$ in steps of 0.1. We place these light curves at a range of redshifts ($0\ \leq\ z\ \leq\ 0.06$). We next measure $t_2(r)$ and $\mathcal{F}_{r_2}$ for all the simulated light curves and find how much the values differ compared to the simulated light curve with the same $x_1$ and $c$ at $z\ =\ 0$ (see Fig.~\ref{kcor}). We adopt the residuals as additional uncertainties due to unaccounted for K-corrections in our measured parameters ($t_2(r/i)$, $\mathcal{F}_{r_2/i_2}$). We select which grid residual point to adopt for each SN based on its redshift, $x_1$, and $c$ in bins of 0.01, 0.5, and 0.1, respectively. The measured $x_1$- and $c$-values are not always reliable for the SNe Ia that are not in our cosmological sample (see Sect. \ref{finalsamples}), but we nonetheless use them here to obtain a rough estimate of the additional uncertainty due to unaccounted for K-corrections.

The variation over the redshift range covered by our sample is $<$ 0.66 d for $t_2(r)$, $<$ 0.20 d for $t_2(i)$, $<$ 0.014 for $\mathcal{F}_{r_2}$, and $<$ 0.087 for $\mathcal{F}_{i_2}$, suggesting that the impact of K-corrections on our results is minimal. As a sanity check, Fig. \ref{kcor} also check how the measured parameters vary for the Hsiao template \citep{Hsiao2007} compared to the SALT3 template with $x_1$ = 0 and $c$ = 0. Although the absolute values of the measured parameters differ from those measured from SALT3, the deviation with redshift is comparable.

\subsection{Outlier rejection}

Although we try to ensure only good quality GP fits are included in our sample by applying a number of cuts detailed in the previous sections, a few poor fits are still present. Due to the large sample size we only inspect the light curves and GP fits for SNe~Ia with extreme $t_2(r)$ or $\mathcal{F}_{r_2}$ values by eye. As a result of these manual inspections, we have removed a few objects from our sample.  

SN 2018but (a normal SN Ia) was removed because the GP function showed unrealistic extrapolation due to a lack of data at $>\ 30$ d. SN 2019sua (91bg-like) was removed because the data between 20--50 d showed spurious fluctuation with underestimated uncertainties, which resulted in unrealistic wiggles in the GP fit. Three additional SNe~Ia were removed (SNe 2018ggw, 2020bxp, and 2020acol, all three normal) because their GP fits produce a dip in the light curve around 20 d even though there is no data in any band. After removing these additional five SNe~Ia, the final sample contains 893 SNe~Ia.

\subsection{Summary of final samples}\label{finalsamples}

We provide a summary of the various cuts applied in Sects. \ref{sample}, \ref{lc_data}, \ref{gp_fits_method}, \ref{second_max_method}, and \ref{second_strength_method}, and the outliers removed in this section that were used to arrive at our main sample and cosmological sample that was presented in Table \ref{cuts_summary}. 

Our final sample consists of 893 SNe Ia. Although the sample is dominated by normal SNe~Ia (706), there are many sub-types of SNe~Ia present. The most ubiquitous are the over-luminous 91T-like SNe~Ia (56), followed by sub-luminous 91bg-like SNe~Ia (36). A full break-down of the various sub-types is shown in Table \ref{table_subtypes}, and a more detailed analysis of the various sub-types and their rates in DR2 will be presented in \cite{Dimitriadis2024}.

In this study we will use the decrease in magnitude between peak and +15 d in the $g$ band, $\Delta m_{15}(g)$, as a measure of the stretch of a light curve. We adopt the $\Delta m_{15}(g)$-values from \cite{Dimitriadis2024}, which are estimated after K-correction using GP fits to the light curves. SALT2 based K-corrections are applied to normal and 91T-like SNe Ia, template K-corrections are applied to all other sub-types, apart from Ia-CSM, which have no applied K-corrections \citep[see][for further discussion]{Dimitriadis2024}. We also aim to perform comparisons between our parameters and those presented in other DR2 cosmology papers, which implement $x_1$ and $c$ measured using SALT2.4 \citep{Guy2007, Taylor2021}. SALT2.4 is trained predominantly on templates of normal SNe~Ia and can therefore not accurately measure the light curve parameters for more peculiar SNe Ia. We therefore define a `cosmology' sub-sample, which is defined using the same cuts as those implemented by \cite{Rigault2024} for the cosmological analysis of the DR2 sample (i.e.$x_1$ and $c$: $|x_1|<3$, $-0.2\leq c\leq0.8$, cuts on the uncertainties of $|\sigma_{x_1}|\leq 1$, $|\sigma_{c}|\leq 0.1$, $|\sigma_{t_0}|\leq 1$, a fit-probability $>\ 1e^{-7}$, and either no sub-classification, or a sub-classification of normal, 91T- or 99aa-like). These cuts result in a sub-sample of 783 SNe~Ia. We will distinguish this sub-sample from our main sample by referring to it as our cosmological sub-sample because it adheres to the cuts normally applied in order to perform cosmological analysis. However, we note that the DR2 is currently not suitable for inferring cosmological parameters \citep[see][for further discussion]{Rigault2024}.

\begin{table}
\centering
 \caption{A summary of the SN Ia sub-types present in our final sample.}\label{table_subtypes}

\begin{tabular}{ c|l } 
\hline \\[-0.8em]
Sub-type & Number \\
\hline \\[-0.8em]
Normal & 706 \\
91T-like & 56 \\
Not sub-typed$^*$ &  50 \\
91bg-like & 36 \\
99aa-like & 19 \\
02cx-like & 13 \\
03fg-like & 7 \\
02es-like$^{**}$ & 4 \\
Ia-CSM & 2 \\
\hline
\end{tabular}
 \begin{flushleft}
    *SNe~Ia which were spectroscopically classified as SNe~Ia but either due to low quality spectra or the phases of the spectra being too late, a sub-classification could not be determined.
    **We include SN 2019yvq \citep{Miller2020b, Siebert2020, Burke2021} in the 02es-like sub-class because it was tentatively classified as a 02es-like SN Ia by \cite{Burke2021}.
 \end{flushleft}

\end{table}

\subsection{Averaged spectral templates}\label{spec_data}

In this paper, we are interested in constraining the spectral properties of SNe Ia from maximum light with the aim of understanding the spectral features contributing before, during, and after the time of secondary maximum (e.g. from peak to +30 d). To investigate this, we have compiled average spectral templates as a function of $x_1$ and phase from a sample of SN Ia spectra from the ZTF DR2 cosmological sub-sample \citep{Johansson2024, Burgaz2024}, supplemented with additional spectra taken by ZTF post-DR2 (obtained between February 2021 -- September 2023) along with literature spectra obtained from \textsc{WISEREP}. The additional spectra were used to increase the sample size at the upper and lower ends of our $x_1$ range at later phases (+15 to +30 d). 

We excluded spectra with a signal-to-noise ratio (S/N) in the continuum of $<$5, and we binned the spectra to 10 \AA\ resolution to account for varying resolution across the sample, in particular the low resolution spectra ($R\ \sim\ 100$) taken by the SEDMachine \citep{Rigault2019, Blagorodnova2018, Kim2022}. In total, we include 265 spectra of SNe Ia in our cosmological sub-sample, where the phase is defined relative to the time of maximum in the $g$ band from the SALT2.4 fit. We also included 9 SN Ia spectra taken by ZTF post-DR2 at $z\leq 0.06$ at phases between +10 and +40 d. We fit the light curves of SNe Ia obtained post-DR2 with SALT2.4 using the same procedure as for the rest of the sample to obtain $x_1$ and $t_0$ values. Finally, we searched \textsc{WISEREP} for spectra taken between +10 and +40 d for normal SNe at $z \leq$0.06 and with published $x_1$ and $t_0$ values, adding a further 17 spectra. These additional spectra were pre-processed in the same way as the ZTF DR2 spectra.

\begin{figure*}
    \centering
    \includegraphics[width=18cm]{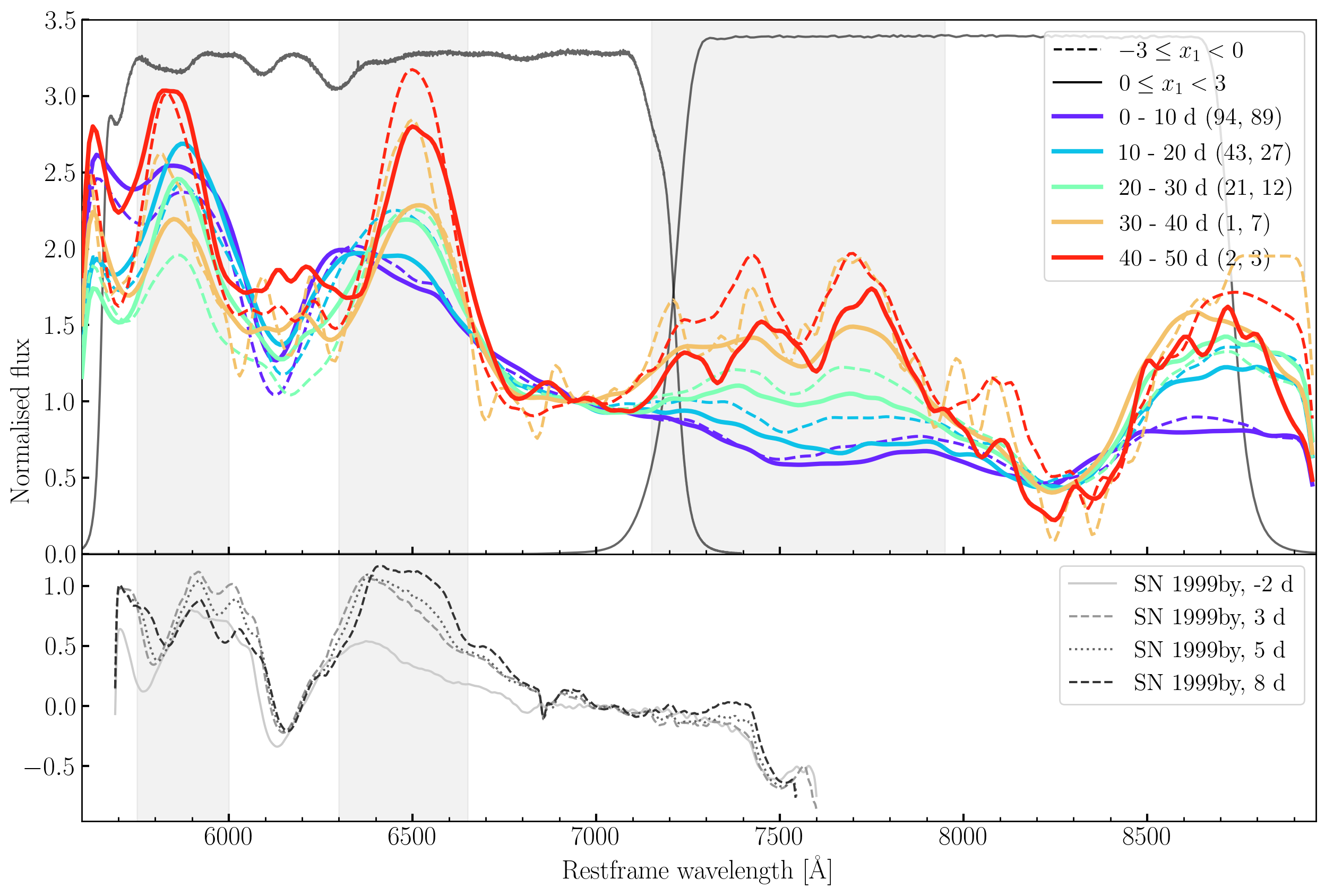}
    \caption{Averaged spectra between 0--10, 10--20, 20--30, 30--40, and 40--50 d in the $r$- and $i$-band wavelength ranges. The phases are relative to $g$-band maximum. Spectra are binned into $-3\leq x_1<0$ (dashed) and $0\leq x_1<3$ (solid). The grey shaded regions indicate the spectral regions where a feature appears during the onset of the secondary maximum. The numbers in the brackets represent the number of spectra going into each bin (first number represents the number of spectra in the $-3\leq x_1<0$ bin, and the second number are the number of spectra in the $0\leq x_1<3$ bin). We also show the filter response functions of the ZTF $r$ and $i$ bands in black. In the bottom panel we show a spectral timeseries of SN 1999by (a 91bg-like SN Ia).}\label{spec_panels_relative_to_max}
\end{figure*}

In Fig. \ref{spec_panels_relative_to_max} we show the $r$- and $i$-band region of the averaged spectra of SNe~Ia from our cosmological sub-sample, sorted into two $x_1$ bins ($-3\leq x_1 <0$ and $0\leq x_1 <3$), and six phase bins (0--10, 10--20, 20--30, 30--40, and 40--50 d). The spectra in each bin were first sigma-clipped (5$\sigma$), then averaged, and finally smoothed with a Gaussian convolution with a standard deviation of 5. All averaged spectra are then normalised to the continuum region at 7000 \AA. We also include spectra of SN 1999by \citep[a 91bg-like SN Ia,][]{Matheson2008} taken from WISEREP in the bottom panel of Fig. \ref{spec_panels_relative_to_max}. The spectra of SN 1999by are also normalised to the continuum region around 7000 \AA.

\section{Results}\label{results}

After the additional cuts on the length scale and uncertainties on $t_2(r)$ and $\mathcal{F}_{r_2}$, as well as the removal of outliers, the main sample remaining consists of 893 SNe~Ia. In this section we test whether the correlations found by \cite{Papadogiannakis2019a} for the CfA and CSP samples persist for our sample. We then analyse the available spectra in order to constrain the spectral feature responsible for the secondary maximum in the $r$ band. We next use the integrated flux under the secondary maximum to estimate the transparency timescales for the SNe~Ia in our main sample, and compare it to model predictions. We also provide estimated total ejected masses derived from the transparency timescale, and compare these to masses derived from the SALT2.4 light curve width parameter, $x_1$, for our cosmological sub-sample.

\subsection{Timing and strength of secondary maximum}\label{timing_strength_results}

\begin{figure*}
    \centering
    \includegraphics[width=17cm]{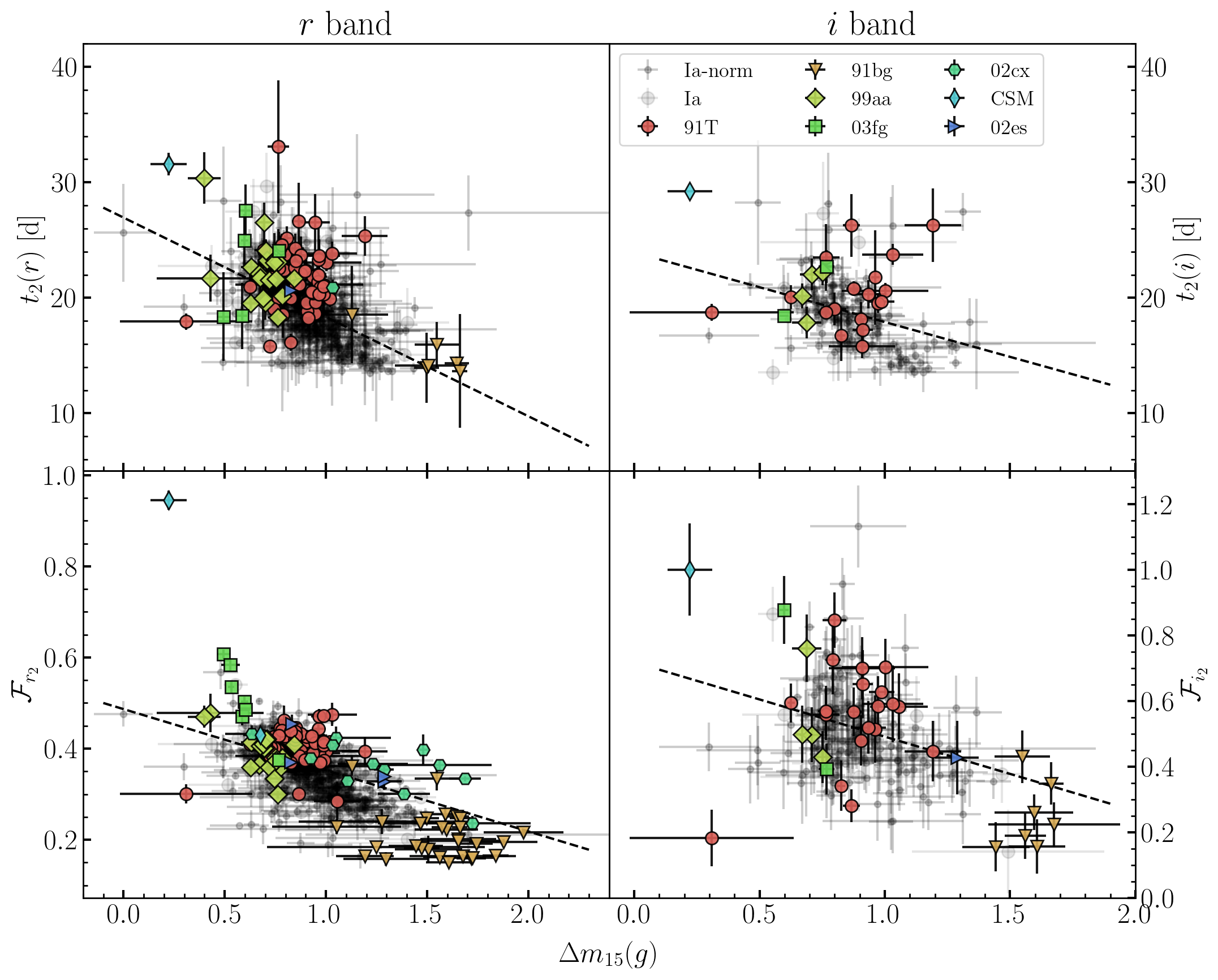}
    \caption{The timing of secondary maximum ($t_2(x)$, top panel) and normalised integrated flux in the range 15--40 d ($\mathcal{F}_{x_2}$, bottom panel) in the $r$ band (\textit{left}) and $i$ band (\textit{right}) as a function of $\Delta m_{15}(g)$ for our sample. We also show simple linear regression fits to the whole sample (black dashed lines). We find a non-zero slope at the $> 5\sigma$ confidence level for $t_2(r)$, $t_2(i)$, $\mathcal{F}_{r_2}$, but not for $\mathcal{F}_{i_2}$ with  $\Delta m_{15}(g)$. }\label{t2_f2_dm15}
\end{figure*}

We have measured the timing and the strength of the secondary maximum for the $r$ and $i$ band. All four parameters, $t_2(r)$, $t_2(i)$, $\mathcal{F}_{r_2}$, and $\mathcal{F}_{i_2}$ are shown as a function of $\Delta m_{15}(g)$ in the four panels of Fig. \ref{t2_f2_dm15}.

\subsubsection{\textit{r} band}

We have measured the timing of the secondary maximum/shoulder in the \textit{r} band, $t_2(r)$, for our ZTF sample and find values ranging between $\sim$13 -- 34 d with respect to maximum light. This is similar to the range seen in \cite{Papadogiannakis2019a} but our values are slightly smaller due to the different definition of $t_2(r)$ (see Sect. \ref{second_max_method}). 

In the top panel of Fig.~\ref{t2_f2_dm15} we show $t_2(r)$ as a function of $\Delta m_{15}(g)$. We find a strong correlation for the normal SN~Ia population, with a Pearson correlation coefficient $r$ of $-0.69$ at a $>5\sigma$ confidence level. 
\cite{Papadogiannakis2019a} found the timing of the secondary maximum to correlate with SNooPy light-curve width parameter, $s_{BV}$, which itself is strongly correlated with $\Delta m_{15}(B)$ \citep{Burns2018}. 
We were able to measure $t_2 (r)$ for a number of unusual classes of SNe~Ia (also shown in Fig.~\ref{t2_f2_dm15}), and find that they broadly follow the normal sample relation. The correlation between $t_2(r)$ and $\Delta m_{15}(g)$ becomes weaker but remains statistically significant when tested for the whole sample rather than just the normal SNe~Ia (Pearson correlation coefficient $r$ of $-0.60$ at a $>5\sigma$ confidence level).

There are seven, out of a total of 46, 91bg-like SNe~Ia with an identified $t_2(r)$, which is unexpected because this sub-class is characterised by a lack of secondary maxima \citep[e.g.][]{Taubenberger2017}. We have inspected all their light curves by eye and for six out of seven, their secondary maxima look real (i.e. there is sufficient data to constrain a secondary maximum and all data points around $t_2(r)$ are consistent with the GP fit, and if $i$-band data is available $t_2(r)$ agrees with $t_2(i)$ within 3$\sigma$). We remove the $t_2$ measurements of SN 2020acbx from the sample because $t_2(i)$ is not coincident with $t_2(r)$ within 3$\sigma$. We further find a secondary maximum for SN 2018jpk, which is a 02cx-like SNe~Ia. The 02cx-like sub-class are not expected to have secondary maxima due to their well-mixed ejecta composition \citep{Jha2017}. We find one SN Ia-CSM \citep[SN 2020uem][]{Sharma2023} with a late secondary maximum ($t_2(r)\ =\ $31.6 d). Although the secondary maximum looks real, it is possible that this increase in flux in the light curve is fuelled by the interaction with the CSM rather than recombination of IGE because the light curve remains approximately flat for $\sim$50 d.

\begin{figure}
    \centering
    \includegraphics[width=8.5cm]{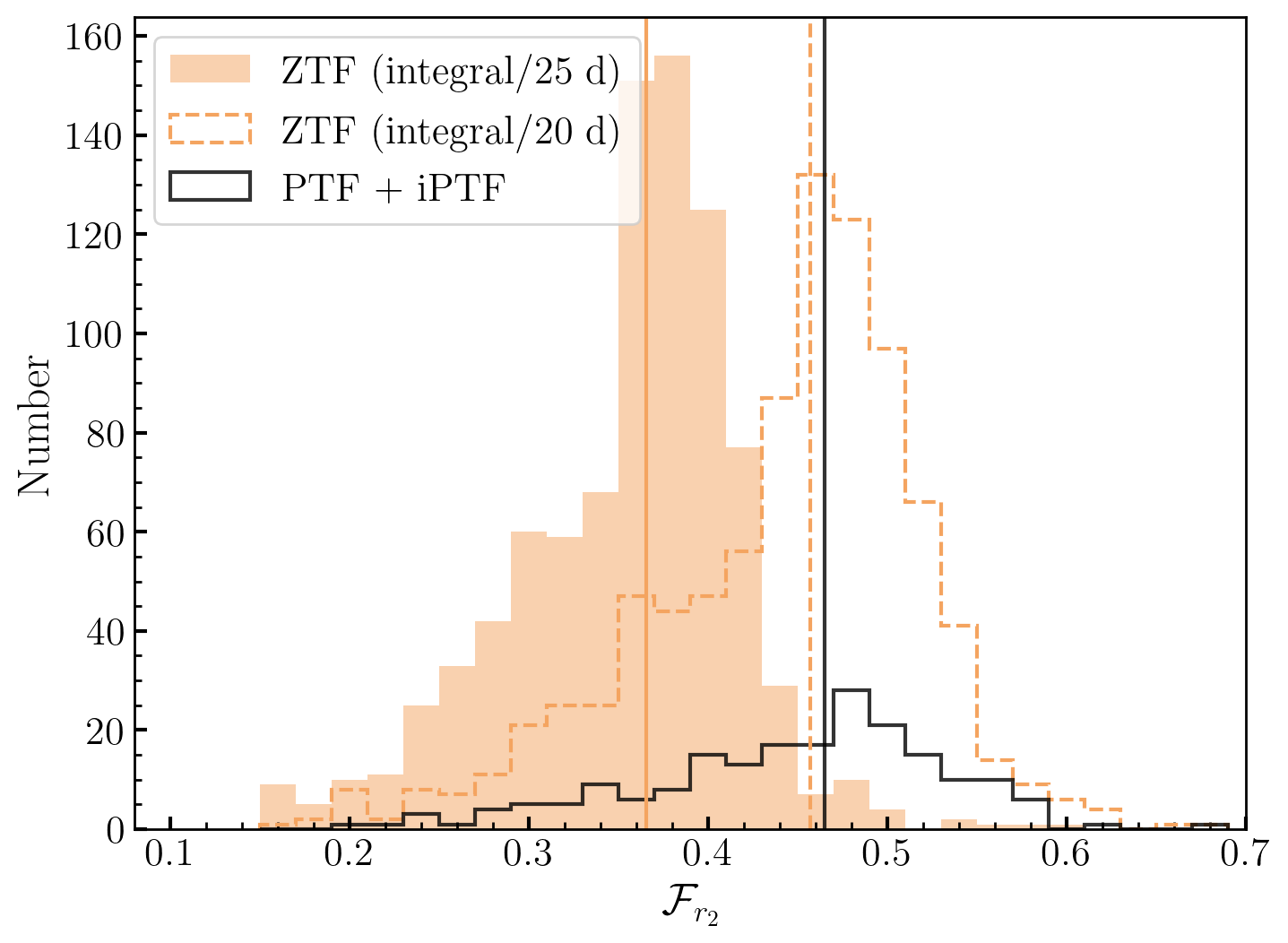}
    \caption{Histogram showing the distribution of $\mathcal{F}_{r_2}$ for the ZTF DR2 sample with the flux integral divided by 25 d (shaded orange) or divided by 20 d (dashed orange), compared to the distribution found for PTF and iPTF by \protect\cite{Papadogiannakis2019a} (black). The median of the PTF+iPTF population is 1.5\protect$\sigma$ higher than the median of the ZTF DR2 population when the integral is divided by 25 d, but the ZTF distribution is consistent with the PTF+iPTF distribution if the integral is divided by 20 d instead.}
    \label{iptf_ztf}
\end{figure}

We measured the normalised integrated flux under the secondary maximum ($\mathcal{F}_{r_2}$), which we use as a measure of the strength of the secondary maximum. The typical range of $\mathcal{F}_{r_2}$ for the normal SNe Ia in our sample is 0.20 -- 0.55, which deviates slightly from what was found by \cite{Papadogiannakis2019a} of 0.25 -- 0.60. In Fig. \ref{iptf_ztf} we compare the distributions of $\mathcal{F}_{r_2}$-values between the two samples and find that the ZTF DR2 population peaks at lower values, with the medians of the two distributions differing by 0.1. We do not expect any intrinsic difference between the SN~Ia samples. However, the integrated flux under the secondary maximum was originally defined as a parameter by \cite{Krisciunas2001} to quantify the strength of the secondary maximum in the $i$ band. \cite{Krisciunas2001} integrated over the range 20--40 d, and normalised by dividing the integral by 20 d. Although \cite{Papadogiannakis2019a} described their integration as over the same range as this paper (15--40 d), if we divide our integrals by 20 d instead of 25 d, the $\mathcal{F}_{r_2}$ distribution from ZTF lines up very well with that of PTF and iPTF (the shifted distribution is also shown in Fig. \ref{iptf_ztf}). As a sanity check, we compare our $\mathcal{F}_{i_2}$-values to those presented in \cite{Krisciunas2001} and they cover the same range, whereas those presented by \cite{Papadogiannakis2019a} are offset to higher $\mathcal{F}_{i_2}$-values, just as their $\mathcal{F}_{r_2}$-values are.

The bottom panel of Fig.~\ref{t2_f2_dm15} shows $\mathcal{F}_{r_2}$ as a function of $\Delta m_{15}(g)$, and we also find a very strong correlation between these parameters for the normal SN Ia population, with a Pearson correlation coefficient $r$ of $-0.72$ at a $>5\sigma$ confidence level. $\mathcal{F}_{r_2}$ is a useful metric because we are able to measure it for all the light curves in our sample, not just those with a measured secondary maxima. This means we are able to include sub-types of SNe~Ia which typically do not show a secondary maximum. When measured for the full sample, the strong correlation remains (a Pearson correlation coefficient $r$ of $-0.71$ at a $>5\sigma$ confidence level).

The most distinct outlier visible in the bottom panel of Fig. \ref{t2_f2_dm15} is a Ia-CSM, which sits 21$\sigma$ above the trend. In general, the 03fg-like sub-class also have larger $\mathcal{F}_{r_2}$ than expected for their $\Delta m_{15}(g)$ value. Since both these sub-classes of SNe~Ia are thought to be powered by some interaction with a dense envelope/CSM \citep{Taubenberger2011, Taubenberger2017, Dimitriadis2021, Dimitriadis2023, Sharma2023}, a higher $\mathcal{F}_{r_2}$ is expected. Some of the 02es-like SNe~Ia also sit above the trend, implying a higher than expected $\mathcal{F}_{r_2}$. \cite{Hoogendam2023} suggested that this sub-class is linked to the 03fg-like sub-class because the two share similar photometric and spectroscopic properties. Interaction with a dense envelope could also explain the higher $\mathcal{F}_{r_2}$ for the 02es-like sub-class. We note that all three of these sub-classes show a range of $\mathcal{F}_{r_2}$-values, with the majority having larger $\mathcal{F}_{r_2}$-values than normal SNe~Ia, but some are consistent with the normal population.

There is a spread in the $\mathcal{F}_{r_2}$ values of the 02cx-like sub-class, with most being consistent with the general trend, but several sitting slightly above the trend. 02cx-like SNe Ia are thought to leave behind a bound remnant which could explain the additional flux at these phases \citep{Kromer2013, Jha2017}. 

The mean $\mathcal{F}_{r_2}$ for the 91T-like sub-class is 0.41, whereas the mean $\mathcal{F}_{r_2}$ for normal SNe Ia in the same $\Delta m_{15}(g)$ range (0.3 $\leq \Delta m_{15}(g)\ \leq$ 1.05) is 0.37. \cite{Leloudas2015} suggested the brighter peak absolute magnitudes of the 91T-like sub-class can be attributed to interaction with a dense CSM originating from a non-degenerate companion. The $\mathcal{F}_{r_2}$-values we find for this sub-class can be interpreted to support this proposition if we assume that the additional flux from the interaction contributes more between 15-40 d than at peak, since $\mathcal{F}_{r_2}$ is calculated from the peak normalised light curves. SN 2020uoo has a weak secondary maximum ($\mathcal{F}_{r_2}\ =\ $0.30), but we note that although the GP fit quality is good, this SN passed our light curve coverage criteria due to having sufficient $i$-band data on the rise, but it has limited $g$- and $r$-band data, which could result in a biased estimate of the time of maximum light and therefore $\Delta m_{15}(g)$. This is unusual since in most cases, ZTF light curves have significantly more data in $g$ and $r$ than $i$.  

It is notable that the 91bg-like SNe~Ia appear like a continuous extension of the normal SN~Ia population in the bottom left panel of Fig. \ref{t2_f2_dm15}, since there is ongoing debate as to whether 91bg-like SNe~Ia have a different physical origin from normal SNe~Ia \citep{Taubenberger2008, Li2022, Graur2023a, Harvey2023}. Based solely on the behaviour of $\mathcal{F}_{r_2}$ as a function of $\Delta m_{15}(g)$, it appears that 91bg-like and normal SNe~Ia are members of the same population.

\subsubsection{\textit{i} band}

\begin{figure}
    \centering
    \includegraphics[width=8.5cm]{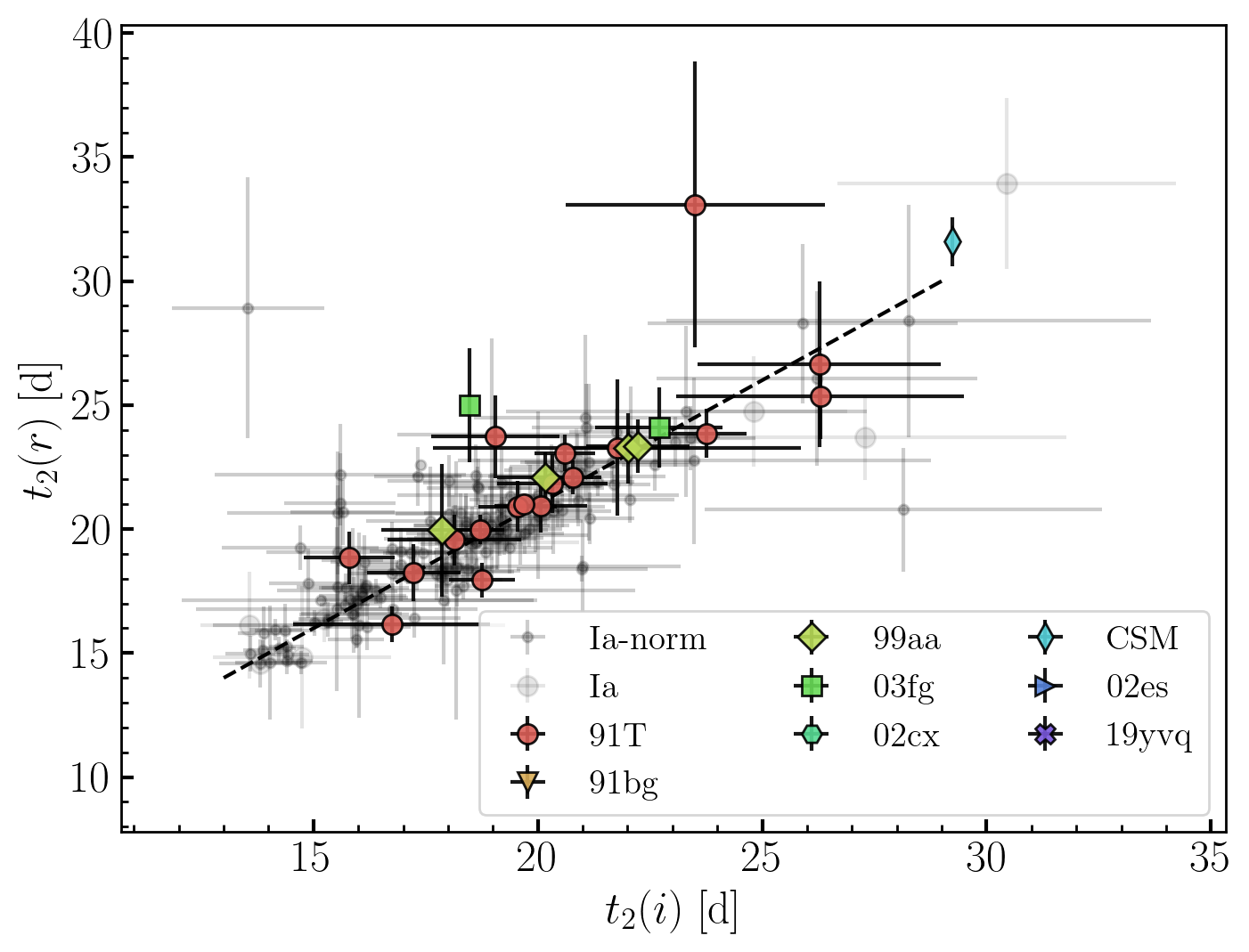}
    \caption{The measured values of $t_2(r)$ vs. $t_2(i)$. We also show a simple linear regression fit (black dashed line) through the full data set.}
    \label{t2_r_vs_i}
\end{figure}

We find that the timing of the secondary maximum is broadly consistent between the $r$ and $i$ band (see Fig. \ref{t2_r_vs_i}), suggesting that the secondary maximum in the two filters share a common cause. The correlation we found between $\Delta m_{15}(g)$ and $t_2(r)$ persists for the $i$ band (see top right panel of Fig.~\ref{t2_f2_dm15}), with a Pearson correlation coefficient $r$ of $-0.45$ at a $>5\sigma$ confidence level for the normal population, and a Pearson correlation coefficient $r$ of $-0.34$ at a $>5\sigma$ confidence level for the full sample. This is slightly weaker than the correlation in the $r$ band, but there are many fewer SNe with sufficient $i$-band data to fit the secondary maximum. 

We also compare $\mathcal{F}_{i_2}$ to $\mathcal{F}_{r_2}$, and find that $\mathcal{F}_{i_2}$ is higher than $\mathcal{F}_{r_2}$ for the majority of the sample (174 out of 182 that could be fitted in the $i$ band), as was found by \cite{Papadogiannakis2019a}. Since the secondary maximum is known to be stronger in the $i$ band than in the $r$ band, a higher integrated flux in this phase range is expected. 

The strong correlation between $\mathcal{F}_{i_2}$ and $\Delta m_{15}(g)$ does not persist for the $i$ band, shown in the bottom right panel of Fig. \ref{t2_f2_dm15}, where we find $r=-0.27$ at a $>5\sigma$ confidence level for the full sample and no significant correlation at all for the normal population alone. The 91T-like SNe~Ia in particular show a wide spread of $\mathcal{F}_{i_2}$ values (0.18 $\leq \mathcal{F}_{i_2} \leq$ 0.85 vs. 0.28 $\leq \mathcal{F}_{r_2} \leq$ 0.47, although we note that the uncertainties are larger in the $i$ band due to the poorer data coverage). Further discussion of the potential causes of the differences in the $r$ and $i$ bands can be found in Sect. \ref{r_vs_i}.

\subsection{Spectral features causing the secondary maximum}\label{identification_spectral_features}

\cite{Jack2015} identify \FeII\ as the dominant spectral feature that causes the secondary maximum in the $i$ band. Their models produce \CoII\ lines and blended \FeII\ features (at 6900 \AA\ and 7500 \AA) in the $r$ band, which strengthen around the time of the secondary maximum. However, these features alone are not sufficient to explain the increase in flux during the $r$-band shoulder. There is another feature around 6500 \AA\ in the spectra of SN 2014J presented by \cite[][see their fig. 8]{Jack2015}, which their model was not able to reproduce.

We can see from Fig. \ref{spec_panels_relative_to_max}, which shows the $r$- and $i$-band wavelength range of averaged spectra of SNe~Ia from our cosmological sub-sample, that there is an emission feature in the $r$ band, red-ward of \SiII\ 6355 \AA\, around 6500 \AA, which become stronger in SNe~Ia with narrower light curves ($-3\leq x_1 <0$) as early as 10--20 d after peak. In SNe~Ia with broader light curves ($0\leq x_1 <3$) this occurs later, around 20--30 d after peak. The flux values are dependent on the region used to normalise the spectra, but the line profiles are similar in the $r$-band region from $\sim$20 d onwards. Since $t_2(r)$ scales with $x_1$ (see Sect. \ref{timing_strength_results}), we suggest that the appearance of these features play a dominant role in the secondary maximum in the $r$ band, just like the \FeII\ feature identified by \cite{Jack2015} causes the secondary maximum in the $i$ band. There is an additional spectral feature blue-ward of \SiII\ 6355 \AA\, around 5850 \AA, which appears to decrease in strength between 10--30 d, but then gets stronger and finally peaks in strength between 40--50 d. However, due to the limited number of spectra in these final bins, and because the spectra of SN 2014J presented by \cite{Jack2015} also showing a decreasing flux in this region, we do not trust that this feature plays a dominant role in the appearance of the secondary maximum. In the $i$ band region of the top panel of Fig. \ref{spec_panels_relative_to_max} we also clearly see the broad \FeII\ feature (between 7200--8000 \AA) which was previously identified by \cite{Jack2015}. This feature increases in strength between 10--40 d, and per each phase bin, it is stronger for spectra of SNe Ia with smaller $x_1$ values.

\begin{figure}
\centering
\begin{subfigure}{0.49\textwidth} 
    \centering
    \includegraphics[width=1\linewidth]{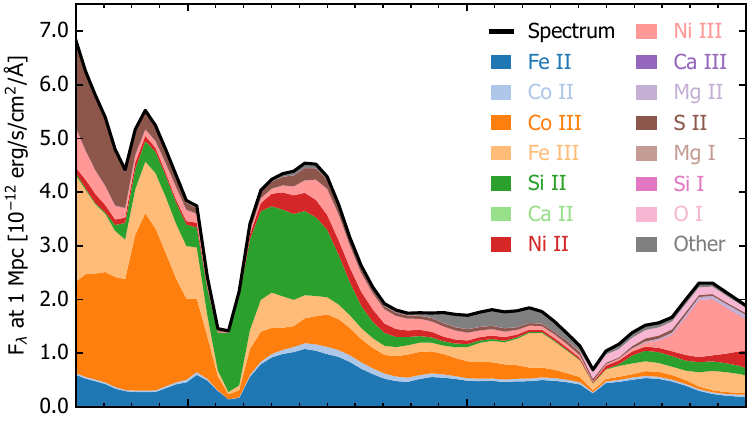}
\end{subfigure}
\begin{subfigure}{0.49\textwidth} 
    \centering
    \includegraphics[width=1\linewidth]{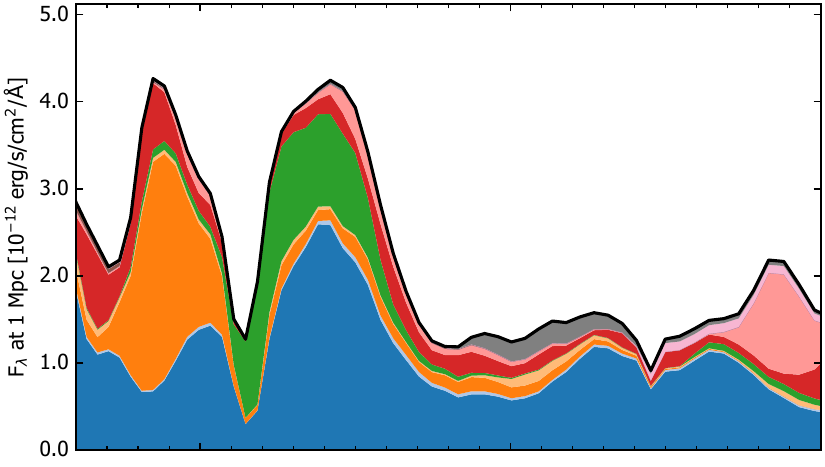} 
\end{subfigure}
\begin{subfigure}{0.49\textwidth} 
    \centering
    \includegraphics[width=1.03\linewidth]{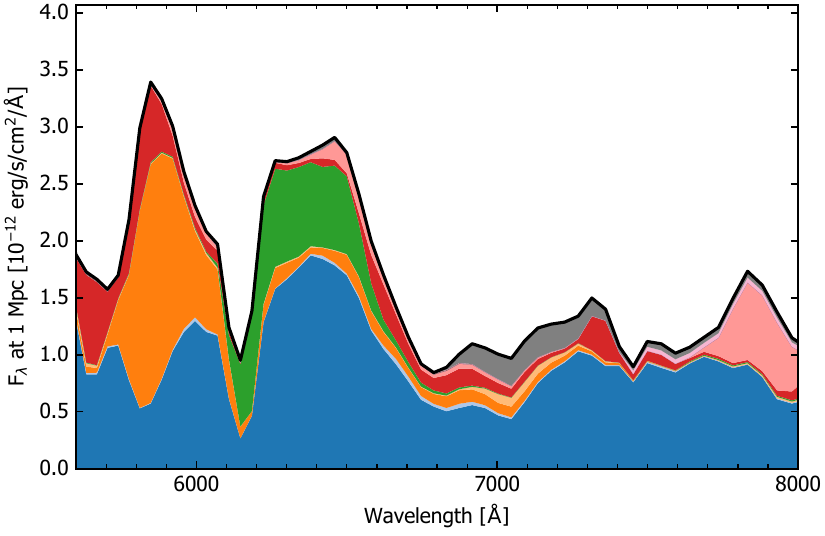} 
\end{subfigure}%
 \caption{Spectra showing the breakdown of species contributing to each feature from a NLTE radiative transfer model of the W7 explosion model at 30 d (top), 40 d (middle) and 50 d (bottom) past explosion.} 
    \label{shingles}
\end{figure}

In Fig. \ref{shingles} we show the breakdown of species contributing to each feature for a non local thermodynamic equilbrium (NLTE) radiative transfer model using \textsc{artis} of a W7 explosion model \citep{Nomoto1984, Shingles2020, Shingles2022} at 30, 40, and 50 d post explosion (approximately 12, 22, and 32 d past maximum light). From this model, it appears that the predominant IGE contributor to the feature at $\sim$6500 \AA\ \citep[which is similar to the unidentified feature in SN 2014J,][]{Jack2015} is \FeII. The feature that sits blue-ward of \SiII\ (around 5850 \AA) has a strong contributions of \CoIII\ and \FeIII\, which fade away as recombination occurs, explaining why this feature initially fades. Although the region also shows a presence of \FeII\, which get stronger as time progresses, the relative strength of the \FeII\ in this spectral feature compared to that at 6500 \AA\ is weaker. The growing contribution of \FeII\ in the $i$ band region between 7200--8000 \AA is also clearly visible in these models. We use the W7 model since it has been shown to be able to reproduce a secondary maximum in the $i$ band and a secondary maximum in the $r$ band \citep{Jack2011}. However, the choice of assumed explosion model can affect the relative contributions of each ion to the net spectrum and therefore, these line identifications are not entirely reliable. 

One of the defining photometric characteristics of 91bg-like SNe~Ia is their lack of secondary maxima \citep{Hoeflich2002, Gonzalez-Gaitan2014, Dhawan2017b, Ashall2020, Hoogendam2022}. It has been argued that these faint events are faster evolving because they have cooler ejecta, enabling recombination to occur sooner, shifting their secondary maximum to earlier times, causing it to blend with the primary maximum \citep{Kasen2006, Blondin2015, Taubenberger2017}. We included spectra of SN 1999by (which is a 91bg-like SN Ia) in the bottom panel of Fig. \ref{spec_panels_relative_to_max}. The feature we have tentatively classified as \FeII, which coincides with the onset of the secondary maximum in normal SNe~Ia, starts to come through in the spectra of SN 1999by as early as $\sim$3 d post maximum light. This is significantly earlier than for the normal SNe~Ia with narrow light curves ($-3\leq x_1 <0$) shown in the top panel of Fig. \ref{spec_panels_relative_to_max}. In the future, we aim to get more detailed spectroscopic time series of 91bg-like SNe~Ia to determine the range of phases at which the recombination occurs for this sub-class.

\subsection{Transparency timescale}\label{transparency_timescale}

\begin{figure}
    \centering
    \includegraphics[width=8.5cm]{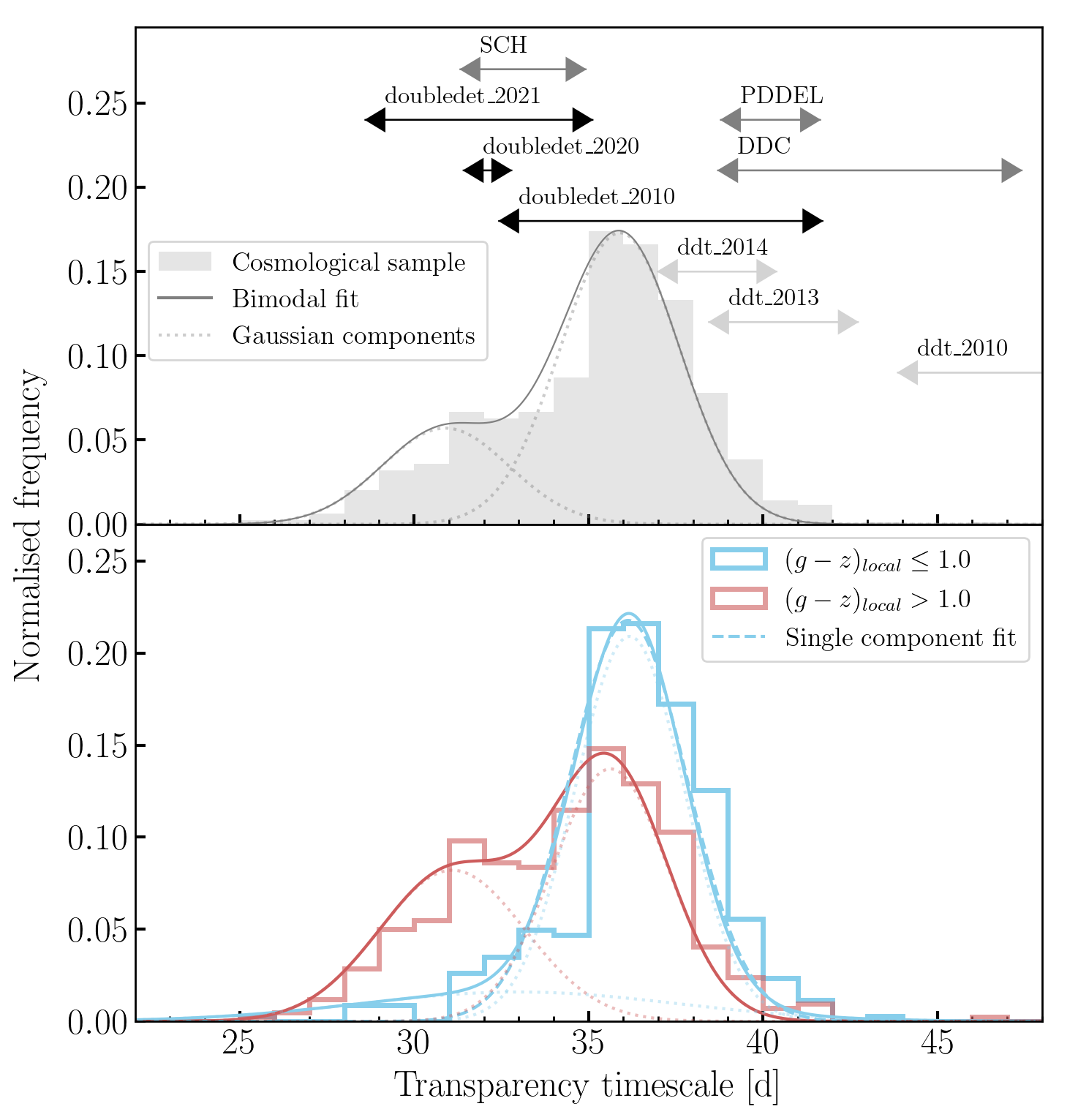}
    \caption{\textbf{Top:} We show the distributions of transparency timescales for the whole cosmological sub-sample (grey), including a bimodal Gaussian fit (solid grey line) and its individual components (dashed grey). The arrows show the ranges of transparency timescales predicted by doubledet\_2010 \citep{Fink2010, Kromer2010}, doubledet\_2020, \citep{Gronow2020}, doubledet\_2021\citep{Gronow2021b}, ddt\_2010 \citep{Fink2010a}, ddt\_2013 \citep{Seitenzahl2013, Sim2013}, ddt\_2014 \citep{Ohlmann2014}, SCH, PDDEL, and DDC \citep{Blondin2017} explosions models. \textbf{Bottom:} We separate the transparency timescale distribution into SNe Ia with blue local colours ($(g-z)_{local}\ \leq \ 1.0$, blue) or red local colours ($(g-z)_{local}\ >\ 1.0$, red). We show bimodal Gaussian the fits to both distributions, and the individual components making up the fit are shown as dotted lines. The Akaike Information Criterion (AIC) test prefers a bimodal fit for the SNe Ia in red local environments, but it does not show a preference for either the double component fit or single component fit for the SNe Ia population in blue environments, so we also show the single component fit as a dashed line to this population. }
    \label{t0_hostmass}
\end{figure}

The transparency timescale is defined by \cite{Jeffery1999} as the epoch at which the ejecta have an optical depth of unity to $\gamma$-rays. For normal SNe~Ia, this value ranges between 30--50 d \citep{Scalzo2014b, Dhawan2017a}. The transparency timescale is a valuable metric because it has been shown to relate directly to the total ejected mass \citep{Jeffery1999, Stritzinger2006a, Scalzo2014b, Dhawan2017a, Dhawan2017b}, which can be used to constrain the explosion models for SNe~Ia.

\cite{Papadogiannakis2019a} were able to calculate pseudo-bolometric light curves for SNe~Ia in their sample provided by CSP because these have multi-wavelength coverage. From the pseudo-bolometric light curves they calculate the transparency timescale ($t_0$). They compare their values of $t_0$ for the CSP sub-sample to their measured $t_2(r)$ and $\mathcal{F}_{r_2}$ values, and find a strong correlation with the latter, providing the following linear relation:

\begin{equation}\label{t0_from_f2_eq}
    t_0 = 44.92 (\pm 5.86) \times \mathcal{F}_{r_2} + 15.00 (\pm 2.32) ,
\end{equation}

\noindent relating $\mathcal{F}_{r_2}$ to $t_0$. Due to the offset in our $\mathcal{F}_{r_2}$ distribution compared to that presented in \cite{Papadogiannakis2019a}, see Sect. \ref{timing_strength_results}, we cannot directly use the above linear correlation to estimate $t_0$ for our sample. Instead, we recalculate $\mathcal{F}_{r_2}$ by dividing the integrated flux between 15--40 d by 20 d (which we will refer to as $\mathcal{\bar{F}}_{r_2}$), ensuring that the distribution matches that of \cite{Papadogiannakis2019a} as shown in Fig. \ref{iptf_ztf}.  

Using Eq. \ref{t0_from_f2_eq} and $\mathcal{\bar{F}}_{r_2}$, we estimate the $t_0$ values for our sample. In the top panel of Fig. \ref{t0_hostmass} we show the distributions of $t_0$ for the cosmological SN Ia sample (see definition in Sect~\ref{finalsamples}). The $t_0$ distribution does not show a single Gaussian distribution but instead appears bimodal, with a stronger peak at longer $t_0$ than the weaker peak at shorter $t_0$ values. This is similar to the bimodal $x_1$ distribution identified in \cite{Nicolas2021} and \cite{Ginolin2024}.

To estimate the properties of this bimodal distribution, we fit the $t_0$ values with a double Gaussian model. The fit is defined as $P(t_0)=r \mathscr{N}(\mu_{high}, \sigma_{high}) + (1-r)\mathscr{N}(\mu_{low}, \sigma_{low})$, where $P(t_0)$ is the probability distribution, $\mathscr{N}$ is the Gaussian distribution, $r$ is the amplitude of the high $t_0$ Gaussian, $\mu$ is the mean and $\sigma$ is the standard deviation of each Gaussian.
We summarise the results of this fit in Table \ref{bimodal_fit}. We find $r\ =\ 0.74\ \pm\ 0.05$, suggesting that the higher $t_0$ mode is the dominant mode over those SNe Ia with lower $t_0$ values. This result is consistent with what was found for the bimodal $x_1$ distribution in \cite{Nicolas2021} ($r\ =\ 0.755\ \pm\ 0.05$), but not with \cite{Ginolin2024}, $r\ =\ 0.60\ \pm\ 0.07$).


\begin{table}[]
    \centering
    \caption{Summary of bimodal fit to the $t_0$ distribution of the cosmological sample discussed in Section \ref{transparency_timescale}.}
    \begin{tabular}{l l}
    \hline \\[-0.8em]
    \vspace{0.5em}
        Parameter &  Result\\
        \hline \\[-0.8em]
        $r$ & $0.74\ \pm\ 0.05$\\
        $\mu_{low}$ & $30.9\ \pm\ 0.4$\\
        $\sigma_{low}$ & $1.8\ \pm\ 0.4$\\
        $\mu_{high}$ & $35.9\ \pm\ 0.1$\\
        $\sigma_{high}$ & $1.7\ \pm\ 0.1$\\
        \hline
    \end{tabular}
    \label{bimodal_fit}
\end{table}

We also compare the distribution of $t_0$ in our sample to theoretical explosion models, akin to fig. 5 in \cite{Papadogiannakis2019a}. We fit the following energy deposition function to the bolometric light curves:

\begin{equation}
\begin{aligned}
    E_{dep} &= \lambda_{Ni} N_{Ni} Q_{Ni, \gamma} \exp{(- \lambda_{Ni} t}) \\
        & +\  \lambda_{Co} \frac{\lambda_{Ni}}{\lambda_{Ni} - \lambda_{Co}} [\exp{(-\lambda_{Co} t}) - \exp{(-\lambda_{Ni} t})] \\
         & \times \ [Q_{Co, \gamma} + Q_{Co, e^+} (1 - \exp{((-t/t_0)^2)})],
\end{aligned}
\end{equation}

\noindent where $\lambda_{Ni}$ and $\lambda_{Co}$ are the inverse of the e-folding times of nickel and cobalt (1/8.8 and 1/111.3 d$^{-1}$, respectively), $N_{Ni}$ represents the number of nickel atoms present, which is calculated from the mass of \nick, which is derived from the peak of the bolometric light curve \citep{Arnett1982}. $Q_{Ni, \gamma}$ and $Q_{Co, \gamma}$ represent the energy released from $\gamma$-ray decays of \nick $\rightarrow $ \cob\ and \cob $\rightarrow$ \fe, respectively (1.75 and 3.61 MeV) and $Q_{Co, e^+}$ is the energy release of the positron decay of \cob $\rightarrow$ \fe\ (0.12 MeV).

We consider two types of explosion models from the Heidelberg Supernova Model Archive \citep[HESMA;][]{Kromer2017}, double detonation and delayed detonation models. Specifically, we include the ranges predicted by doubledet\_2010 \citep{Fink2010, Kromer2010}, doubledet\_2020, \citep{Gronow2020}, doubledet\_2021\footnote{The models presented in \cite{Gronow2021b} extend the range of core and He shell masses and were produced to explain the variations seen in the bolometric magnitudes of SNe Ia.} \citep{Gronow2021b}, ddt\_2010 \citep{Fink2010a}, ddt\_2013 \citep{Seitenzahl2013, Sim2013}, and ddt\_2014 \citep{Ohlmann2014}. The $t_0$ ranges predicted by these models are shown in the top panel of Fig. \ref{t0_hostmass}. We also include model predictions from \cite{Blondin2017} for pure central detonations of sub-M$_{ch}$ WDs (SCH), pulsational M$_{ch}$ delayed-detonations (PDDEL), and standard M$_{ch}$ delayed-detonations (DDC). We checked the $t_0$ distributions for gravitationally confined detonation \citep{Seitenzahl2016, Lach2022} and merger models \citep{Pakmor2010, Pakmor2012, Kromer2013a, Kromer2016a}, but the shortest $t_0$ value produced by these models was 45 d and their $t_0$ values extended out to 70 d, which is not consistent with any of our observations. 

We find that the $t_0$ predictions from the doubledet models are consistent with 97 per cent of our cosmological sub-sample, whereas only 28 per cent are consistent with the predictions from the ddt models. From the models presented by \cite{Blondin2017}, the $t_0$ values predicted by the SCH models are consistent with the largest percentage of our sample (26 per cent). The other two explosion models, DDC and PDDEL, are consistent with equal percentages of the sample (8 per cent). 

We also check which explosion model $t_0$ predictions are most consistent with the common SN Ia sub-types. The 91bg-like sub-class spans 22.2 $\leq \ t_0 \leq$ 35.3 d with a median of 26.1 d, and the 91T-like sub-class spans 31.0 $\leq \ t_0 \leq$ 41.7 d with a median of 38.1 d. None of the 91bg-like SNe Ia are consistent with the predictions from the ddt explosion models, 18 per cent are consistent with the doubledet explosion models, and only 3 per cent are consistent with the SCH explosion models. On the other hand, 100 per cent of the transparency timescales of the 91T-like SN Ia population are consistent with the doubledet, 76 percent are consistent with ddt, 40 per cent are consistent with both PDDEL and DDC, and only 3 per cent are consistent with the SCH explosion models. 

In order to test whether the low $t_0$ component originates from SNe Ia in particular host galaxy environments, we bin the distribution into locally blue or red environments ($(g-z)_{local}\ \leq \ 1.0$ or $(g-z)_{local}\ >\ 1.0$, shown in the bottom panel of Fig. \ref{t0_hostmass}). We fit the resulting two distributions with either a single or bimodal Gaussian, the only priors being that the amplitudes of the Gaussians cannot be negative and they cannot share the same mean. We use the Akaike Information Criterion \citep[AIC,][]{Burnham2004}, which is a good test for comparing models because it penalises for additional parameters, to determine whether the population is better described by a single or double component Gaussian. We find that the SNe Ia in locally blue environments can be fit by either a single or double component, with the AIC values differing by $<2$, which is the standard cut-off point for determining whether one model is preferred over the other \citep{Burnham2004}. The distribution of SNe Ia in locally red environments on the other hand shows a strong preference for two components, with an AIC value 39 units lower than for the single component fit.

\subsection{Total ejected mass}

One of the strongest constraints for different progenitor scenarios is the total ejected mass ($M_{\textnormal{ej}}$), since most explosion scenarios can be placed in one of two categories: $M_{ch}$ or sub-$M_{ch}$. Unfortunately, we cannot calculate the total ejected mass directly since this requires bolometric light curves \citep[e.g.][]{Arnett1982, Jeffery1999}, which are not available for the ZTF sample where we only have $g$, $r$, and $i$ coverage. 

\begin{figure}
    \centering
    \includegraphics[width=8.5cm]{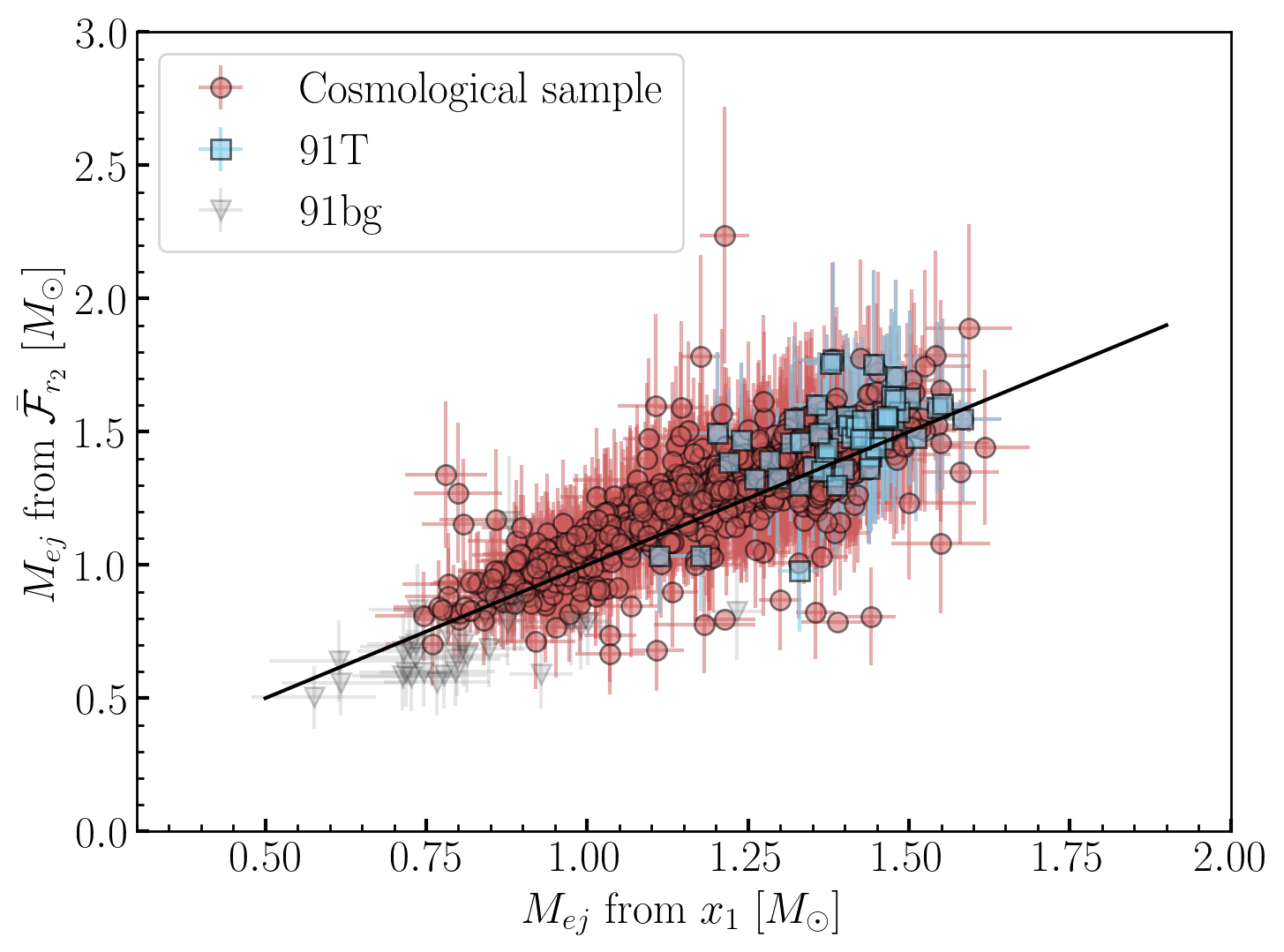}
    \caption{A comparison between the total ejected mass derived from $\mathcal{\bar{F}}_{r_2}$ and $x_1$ for the cosmology sub-sample (which contains the 91T-like SNe Ia) as well as the 91bg-like SN Ia population. The black line is the line of equality. The median total ejected mass calculated from $\mathcal{\bar{F}}_{r_2}$ is $1.3\ \pm \ 0.3$ \msun\ and the median when calculated from $x_1$ is $1.2\ \pm \ 0.2$ \msun. We find a range of 0.64 -- 1.91 \msun\ for $M_{\textnormal{ej}}$ for the normal SNe Ia in the cosmological sample, ignoring the single outlier at 2.24 \msun.}
    \label{ejectedmass_distributions}
\end{figure}

However, we can use our $t_0$ values (which were derived from $\mathcal{\bar{F}}_{r_2}$) to estimate $M_{\textnormal{ej}}$ using the equation; 

\begin{equation}\label{ejectas_mass_eq_t0}
    M_{\textnormal{ej}} / \textnormal{M}_{\odot} = 1.38 \cdot \left( \frac{1/3}{q}\right) \cdot \left( \frac{v_e}{3000\ \textnormal{km s}^{-1}}\right)^2 \cdot \left( \frac{t_0}{36.80 \ \textnormal{d}}\right)^2 
\end{equation}

\noindent which was presented in this form by \cite{Papadogiannakis2019a} but originally derived by \cite{Jeffery1999}. In Eq. \ref{ejectas_mass_eq_t0}, $q$ is a qualitative description of the distribution of the material of the ejecta (a high value for $q$ implies a more centrally concentrated ejecta) and $v_e$ is the e-folding velocity assuming an exponential density profile. We adopt the same values as \cite{Papadogiannakis2019a}: $q\ =\ 1/3$ and $v_2\ =\ 3000$ km s$^{-1}$, leaving $t_0$ as the only variable. These assumed values are approximately valid for normal SNe~Ia, with more or less luminous SNe~Ia having higher or lower e-folding velocities, respectively. We calculate $M_{\textnormal{ej}}$ for the SNe~Ia in our cosmological sub-sample using the $t_0$ values calculated in Sect. \ref{transparency_timescale} (see Fig. \ref{ejectedmass_distributions}). Although they are a member of the cosmological sub-sample, we highlight the 91T-like SN Ia population in Fig. \ref{ejectedmass_distributions}, which sit at the upper end of the $M_{\textnormal{ej}}$ distribution. We also present the calculated $M_{\textnormal{ej}}$-values for the 91bg-like population in Fig. \ref{ejectedmass_distributions}, but because the assumptions on $q$ and $v_e$ are not necessarily valid for this sub-populations, their measured ejecta masses may be less accurate. 

\cite{Scalzo2014b} studied SNe~Ia with multi-wavelength coverage (enabling the construction of bolometric light curves) and measured the total ejected mass using a comparable method to \cite{Papadogiannakis2019a} by measuring $t_0$ from the bolometric light curve by fitting a radioactive decay energy deposition curve to the tail of the observations. However, unlike \cite{Papadogiannakis2019a}, \cite{Scalzo2014b} infer the ejected mass in a Bayesian context using a semi-analytic model of the ejecta. \cite{Scalzo2014b} found a strong correlation between the $M_{\textnormal{ej}}$ and $x_1$, providing the following equation;

\begin{equation}
    M_{\textnormal{ej}}/ \textnormal{M}_{\odot} = (1.253 \pm 0.022) + (0.172 \pm 0.021)x_1 \label{x1_eq}
\end{equation}

\noindent which we use as a secondary method to derive $M_{\textnormal{ej}}$. We note that the $M_{\textnormal{ej}}$-values derived using this equation are only reliable for our cosmological sub-sample of the SNe~Ia since they have reliable $x_1$ values (see Sect. \ref{finalsamples}), but we do also show the 91bg-like SN Ia population for comparison in Fig. \ref{ejectedmass_distributions}. 

Although in general we find a slightly higher ejected mass from $\mathcal{\bar{F}}_{r_2}$ than $x_1$ (see Fig. \ref{ejectedmass_distributions}), the median ejected masses derived from the two methods are consistent with each other (with a median calculated from $\mathcal{\bar{F}}_{r_2}$ of $1.3\ \pm \ 0.3$ \msun\ and a median calculated from $x_1$ of $1.2\ \pm\ 0.2$ \msun). This is despite the fact that the correlation used to derive $t_0$ from $\mathcal{\bar{F}}_{r_2}$ presented by \cite{Papadogiannakis2019a} is based on 17 SNe~Ia and includes two faint 91bg-like SNe~Ia, whereas the correlation presented by \cite{Scalzo2014b} is based on 21 SNe~Ia and includes normal and bright SNe~Ia, such as 91T-like or 03fg-like SNe~Ia. These two studies are sampling opposite extremes of the absolute magnitude distribution of SNe~Ia, and it has not been established whether these extremes are members of the same population as normal SNe~Ia. Although 91T-like SNe~Ia have very similar properties to normal SNe~Ia, studies have suggested that both 91bg- and 03fg-like SNe~Ia could originate from different progenitor scenarios or explosion mechanisms than normal SNe~Ia \citep{Stritzinger2006, Taubenberger2008, Taubenberger2013, Scalzo2014b, Taubenberger2019, Graur2023, Hoogendam2023}. Moreover, the method implemented by \cite{Papadogiannakis2019a} to derive $M_{\textnormal{ej}}$ from $t_0$ differs from that implemented by \cite{Scalzo2014b}, with \cite{Scalzo2014b} using a Bayesian framework. However, since both methods use the bolometric light curves to derive $t_0$, it is not surprising that the resulting distributions are consistent with each other, but it does suggest that the correlation used to derive $t_0$ from $\mathcal{F}_{r_2}$ provides reliable results.

\section{Discussion}\label{discussion}

In this section we discuss the differences between the $r$ and $i$ band secondary maxima. We also analyse a potential non-linearity we find in the correlation between $x_1$ and $\mathcal{F}_{r_2}$, and compare this to recent literature results which find a similar trend with Hubble residuals. We briefly discuss the potential of using the secondary maximum in the $r$ band for standardisation. We introduce \textsc{TURTLS} radiative transfer model results that enable us to estimate the total mass of \nick\ produced in the explosion, and use this in combination with the results presented in the previous section to try to constrain the dominant parameter affecting the secondary maximum in the $r$ band.

\subsection{Differences between the \protect$r$ and \protect$i$ band secondary maximum}\label{r_vs_i}

In Sect. \ref{timing_strength_results} we presented the strength and timing of the secondary maximum in the $r$ and $i$ bands. Whilst we found the timing of the secondary maximum to be consistent between the $r$ and $i$ bands, the integrated flux under the secondary maximum is larger in the $i$ band. Furthermore, we find strong correlations between $\Delta m_{15}(g)$ and $t_2(r)$, $t_2(i)$, and $\mathcal{F}_{r_2}$ but not with $\mathcal{F}_{i_2}$.

The secondary maximum is known to be more prominent in the $i$ band than in the $r$ band \citep{Kasen2006}. The integrated flux is normalised to peak, but because the secondary maximum is stronger in the $i$ band relative to peak, it is not surprising that we find $\mathcal{F}_{i_2}$ to be higher than $\mathcal{F}_{r_2}$ in general.

It is notable that the strong correlation seen between $\mathcal{F}_{r_2}$ and $\Delta m_{15}(g)$ does not persist for the $i$ band. The strong correlation in the $r$ band can be attributed to the nickel mass being one of the main drivers of the strength of the secondary maximum \citep{Kasen2006}. Whilst the integrated flux under the secondary maximum is normalised to peak and should not show a trend with the peak magnitude \citep[driven by the nickelmass,][]{Arnett1982}, a higher mass of nickel also leads to a higher temperature. This higher temperature can result in a higher opacity of the ejecta that increases the timescale of the evolution of the light curve \citep{Mazzali2001, Hoflich1996}. Therefore, if more nickel is produced in the explosion, we expect a smaller $\Delta m_{15}(g)$ and a more prominent secondary maximum, which is what we see in the $r$ band. 

The additional scatter seen in the $i$ band for $\mathcal{F}_{i_2}$ compared to in the \textit{r} band could imply that there is another factor influencing the strength of the secondary maximum, which has a larger effect in the $i$ band than in the $r$ band, and does not correlate with the decline rate. \cite{Kasen2006} found that changing the metallicity affects the secondary maximum whilst having little to no impact on the primary maximum. This could explain the spread seen in the strengths of the secondary maxima in the $i$ band for a constant decline rate in the bottom right panel of Fig. \ref{t2_f2_dm15}. We speculate that the reason this effect is more visible in the $i$ band than in the $r$ band is because it impacts the strength of the secondary maximum but not the decline rate post-maximum. With the secondary maximum being weaker in the $r$ band, the overall decline rate will make up a larger contribution of $\mathcal{F}_{r_2}$ than for $\mathcal{F}_{i_2}$, where the flux in the secondary bump dominates.

\subsection{Non-linear correlation between \protect$x_1$ and \protect$\mathcal{F}_{r_2}$}

In Sect. \ref{timing_strength_results} we found a strong correlation between $\mathcal{F}_{r_2}$ and $\Delta m_{15}(g)$, supporting the findings of \cite{Papadogiannakis2019a}. Here we test whether this correlation extends to $\mathcal{F}_{r_2}$ with $x_1$. We opted to use $\Delta m_{15}(g)$ for the majority of this study because it can be measured from our GP fits, enabling us to measure it for all sub-types of SNe~Ia, whereas $x_1$ relies on a reliable SALT2.4 fit, and since SALT2.4 templates are predominantly trained on normal SNe~Ia it breaks down for more peculiar sub-types. This means that we can only reliably use $x_1$ values for our cosmological sub-sample.

\begin{figure}
    \centering
    \includegraphics[width=8.5cm]{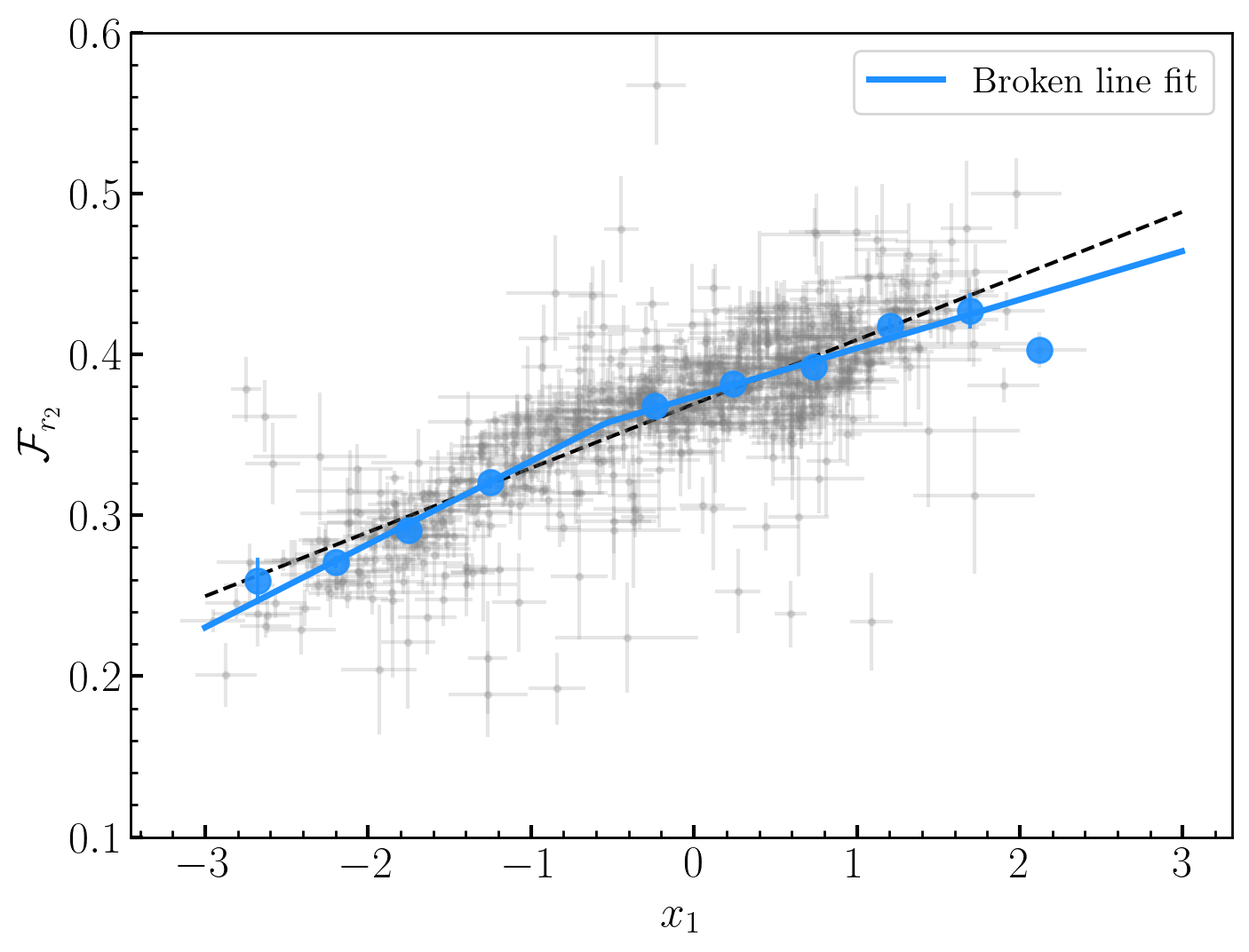}
    \caption{$\mathcal{F}_{r_2}$ as a function of $x_1$ for SNe~Ia in our cosmological sub-sample. We fit the population with a broken line (blue) and linear fit (black, dashed). We find the best-fit value of the stretch split to be at $x_1^0\ =\ -0.5\ \pm\ 0.2$. The blue points show the data binned by stretch, taking the mean and standard error of the mean in each bin. }
    \label{x1_f2_brokenline}
\end{figure}

We plot $\mathcal{F}_{r_2}$ as a function of $x_1$ for the cosmological sub-sample in Fig. \ref{x1_f2_brokenline} and as for $\mathcal{F}_{r_2}$ and $\Delta m_{15}(g)$, we find a $>5\sigma$ correlation between these parameters. \cite{Ginolin2024} find a strongly non-linear stretch-magnitude relationship (a change of slope at $x_1^0$, with $x_1^0\ =\ -0.53\ \pm\ 0.09$). \cite{Senzel2024} find a similar result with the local host colour and $x_1$. Motivated by these studies, we test for similar non-linearity between $\mathcal{F}_{r_2}$ and $x_1$ by fitting the sample with a `broken-line':

\begin{equation}
    \mathcal{F}_{r_2} = \begin{cases}
          m_1 x_1 + c_1 \quad &\text{if} \,\ x_1 \leq x_1^0 \\
          m_2 x_1 + c_2 \quad &\text{if} \,\ x_1>x_1^0 \\
     \end{cases}
\end{equation}

\noindent where $x_1^0$ is the $x_1$ value at which the line breaks and this is a free parameter which is allowed the vary over the full $x_1$ range of the sample, $m_1$ and $m_2$ are the slopes and $c_1$ and $c_2$ are the y-intercepts before and after the line break ($x_1^0$), respectively. We find a split in the population at$x_1^0\ =\ -0.5\ \pm\ 0.2$, and the broken-line fit is also shown in Fig. \ref{x1_f2_brokenline}. Using AIC as a test to compare the models, the `broken-line' model is statistically preferred (difference of 52 AIC units). Interestingly, the best-fit value of the stretch split ($x_1^0\ =\ -0.5\ \pm\ 0.2$) is consistent with what was found by \cite{Ginolin2024}, ($x_1^0=-0.49\ \pm\ 0.06$) where they argue that the $\alpha$ term in SN~Ia standardisation is non-linear, with the low stretch mode only originating from older populations and the high stretch mode containing SNe~Ia from both old and young populations \citep{Nicolas2021}.

\subsection{The secondary maximum in the \protect$r$ band for standardisation}

\begin{figure*}
\centering
    \includegraphics[width=18cm]{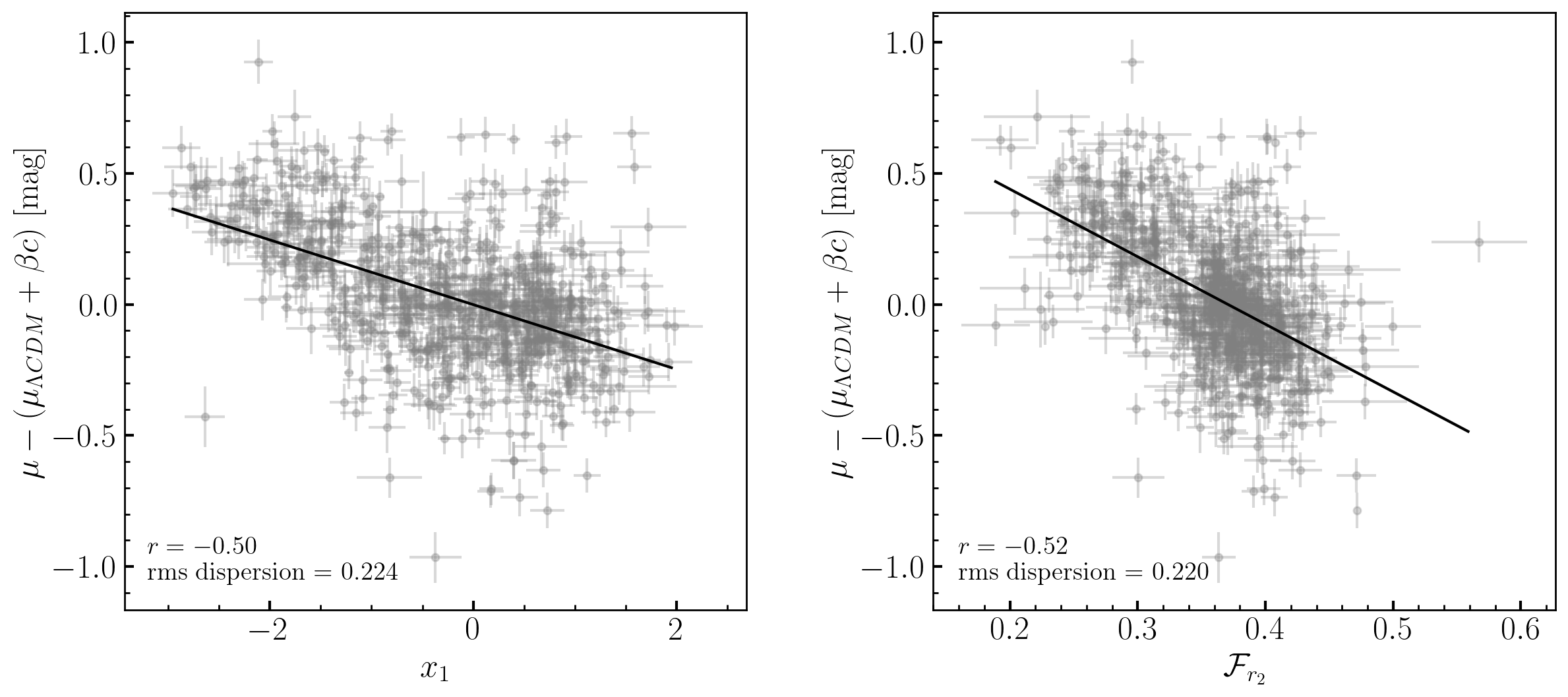}
    \caption{Colour corrected Hubble residuals as a function of $x_1$ (left) and $\mathcal{F}_{r_2}$ (right) for the cosmological sub-sample. Pearson's $r$ correlation coefficients and the rms dispersion of the Hubble residuals are shown in the plots. }\label{standardisation_plots}
\end{figure*}

We also test the secondary maximum in the $r$ band as an alternative standardisation parameter in the absence of $x_1$, as \cite{Shariff2016} did for the secondary maximum in the $J$ band. To this end we use colour-corrected Hubble residuals \citep[not corrected for $x_1$ or the host mass step, values adopted from][]{Ginolin2024}, and compare the correlation with $x_1$, $t_2(r)$, and $\mathcal{F}_{r_2}$. We find that although the correlation with $t_2(r)$ is statistically significant ($>5\sigma$), the correlation between the Hubble residual and $t_2(r)$ is weaker than with $x_1$, with Pearson's correlation coefficients of $r=-$0.34, $-$0.50, respectively. $\mathcal{F}_{r_2}$ on the other hand shows a slightly stronger correlation with the colour-corrected Hubble residuals, with $r=-0.52$ at the $>5\sigma$ confidence level. The root mean square (rms) dispersion of the Hubble residuals is also reduced when using $\mathcal{F}_{r_2}$ instead of $x_1$ (rms = 0.220 and 0.224 mag, respectively). We perform bootstrapping to estimate the robustness of these results \citep{Efron1983, Efron1997}. We sample data points with replacement from the original sample 1000 times to generate of alternative sub-samples of the same size. We compute 95 per cent confidence intervals from our bootstrapping and find $-0.58 \leq r \leq -0.46$ for the correlation between the colour-corrected Hubble residuals and $\mathcal{F}_{r_2}$. The 95 per cent confidence interval for the correlation with $x_1$ is $-0.55 \leq r \leq -0.42$, meaning these correlations are consistent. The correlations between the colour-corrected Hubble residuals and $x_1$ as well as $\mathcal{F}_{r_2}$ are shown in Fig. \ref{standardisation_plots}.

\cite{Hayden2019} found that $x_1$ measured from the rising part of the light curve showed a stronger correlation with the peak magnitude than $x_1$ calculated from the post-peak light curve. Our results are not aligned with this and suggest that the post-peak light curve contains more information. Although it is interesting that we find $\mathcal{F}_{r_2}$ to be a better standardisation parameter than $x_1$, the measurement of $\mathcal{F}_{r_2}$ relies on the presence of data at phases $>10$ d, at which time the SN will have dimmed significantly compared to at maximum light. Moreover, the measurement of $\mathcal{F}_{r_2}$ also requires data around maximum light because the peak magnitude in the $r$ band is required to normalise the light curve to peak. Consequently, standardisation using $x_1$ requires less data and would be preferable in most cases. 

\subsection{Which intrinsic parameters impact the secondary maximum?}

The timing of the secondary maximum is thought to be a temperature effect: the lower the temperature, the earlier the shift in the dominant ionisation state of the IGEs. However, the models presented by \cite{Kasen2006} showed that the amount of stable and radioactive IGEs produced in the explosion, the distribution of IGEs throughout the ejecta, and metallicity may also affect the secondary maximum. We will use our observational data to test the impact of these parameters on the secondary maximum in the $r$ band.

We first investigate the impact of the temperature of the ejecta on the timing of the secondary maximum by checking if there is evolution of $t_2(r)$ with the pseudo-equivalent width (pEW) of the \SiII\ features, using the pEW values presented by \cite{Burgaz2024}. Specifically, we use the pEW of the 5972 \AA\ line as a proxy for the temperature of the ejecta. The ratio between the pEWs of the 5972 and 6355 \AA\ lines is more commonly used as a proxy for temperature \citep{Nugent1995, Hachinger2008}, but \cite{Burgaz2024} find that the pEW of the 5972 \AA\ line is a better tracer of peak magnitude/$x_1$ and therefore temperature, because the 6355 \AA\ line becomes saturated for faint SNe Ia. Our findings agree with this, since we find a statistically significant linear trend between $t_2(r)$ and the pEW of the 5972 \AA\ (Pearson's correlation coefficient $r=-0.38$ at the $>5\sigma$ level), but no statistically significant trend between $t_2(r)$ and the ratio of the pEWS of the 5972 \AA\ and 6355 \AA\ lines (2.5$\sigma$). Assuming the pEW of the 5972 \AA\ \SiII\ line is a good proxy for temperature, this correlation indicates that $t_2(r)$ is impacted by temperature, as expected. However, the correlation is weak (Pearson's correlation coefficient $r=-0.38$). This could be driven by the fact that SNe Ia with the largest $t_2(r)$ values are expected to have very weak \SiII\ 5972 \AA\ pEW values, and in some cases may not be measurable. At the other extreme, SNe Ia with strong \SiII\ 5972 \AA\ pEW values may not have a measurable secondary maximum. If both of these extremes were measurable, it could strengthen the correlation. Alternatively, there could be other factors affecting the timing of the secondary maximum, such as nickel-mixing, metallicity, or nickel mass, although we note that the nickel mass should also strongly correlate with temperature.

In order to test the impact of the mass of IGEs produced and their distribution throughout the ejecta, we fit the $g$- and $r$-band light curves in our sample with the \textsc{TURTLS} 1D radiative transfer code, which models the early light curves of thermonuclear SNe \citep{Magee2018, Magee2020}. We use the same suite of 355 $M_{ch}$-models as presented in \cite{Deckers2022}, which contains four ejected \nick-masses (0.4, 0.5, 0.6, and 0.8 \msun), five different \nick-distributions, two density profiles, and nine different kinetic energies of the ejecta. We follow the fitting method described in \cite{Deckers2022}. In order to reliably constrain the best fitting model, the light curve needs to have good data coverage, especially at early times \citep{Magee2020}. We require the first data point to be no later than 14 d before maximum light in the \textit{g} band, and a minimum cadence of 3 d. This reduces the number of light curves that we are able to fit with the \textsc{TURTLS} models to 499. We take the \nick-mass and \nick-distribution from the best matching \textsc{TURTLS} model to each SN Ia, and the 3$\sigma$ confidence interval is determined by the range of model that fall within 3$\sigma$ of the normalised reduced-$\chi^2=1$ \citep[see][for further details]{Deckers2022}. Similarly to the findings in \cite{Deckers2022}, we find that the majority of the SN Ia light curves are fit with the two most extended nickel distributions and are not able to test for correlations with this parameter. We are able to analyse the effect of nickel mass and find that SNe~Ia with a higher estimated nickel mass tend to have a later secondary maximum at the 3.6$\sigma$ level, as well as a stronger secondary maximum at the 2.5$\sigma$ level, which agrees with the prediction made by \cite{Kasen2006}. We note that the nickel mass parameter in the \textsc{TURTLS} models measures the amount of radioactive nickel produced in the explosion, whereas the study by \cite{Kasen2006} considered the total mass of IGE, combining radioactive and stable material.

\cite{Kasen2006} also showed that metallicity impacts the secondary maximum, but we are unable to measure metallicity for the SNe~Ia in our sample. However, \cite{Kasen2006} predicted that as metallicity increases, the secondary maximum is pushed to earlier phases and that earlier secondary maxima should be as bright, if not brighter than later ones. Our observations of normal SNe~Ia in the $r$ band do not support this because $\mathcal{F}_{r_2}$ shows a positive linear trend with $t_2(r)$ at a $>5\sigma$ confidence level, meaning that in our sample, later secondary maxima tend to be brighter. However, we do not find a significant positive trend between $\mathcal{F}_{i_2}$ and $t_2(i)$. As discussed in Sect. \ref{r_vs_i}, one interpretation of these results could be that a change in metallicity has a stronger impact on the secondary maximum in the $i$ band than in the $r$ band. 

The models presented in \cite{Kasen2006} predict a third bump in SN Ia light curves at around $\sim$ 80 d post maximum light in the $B$ band, due to the recombination of singly ionised iron lines to neutral iron. Unfortunately, the ZTF DR2 light curves do not have sufficiently high S/N at these phases to confirm or reject this prediction. Moreover, \cite{Kasen2006} states that the LTE approximations made to produce the models are not applicable at these late epochs, and this may result in a further delay of the third maximum as well as a weakening of the bump. \cite{Kasen2006} also predicted that the third bump should be most prominent in the $J$ band, but the model predictions presented by \citep[][appendix C]{Blondin2015} show that the only strong \FeI\ features at 100 d post explosion exist in the $U$ band, and so it is unlikely that this recombination would leave any visible signature in the $r$/$i$-band photometry. However, we would encourage a future study to focus on getting high S/N data, in a range of wavelengths, at epochs between 80--150 d to search for signatures of the third maximum.


\begin{figure}
    \centering
    \includegraphics[width=8.5cm]{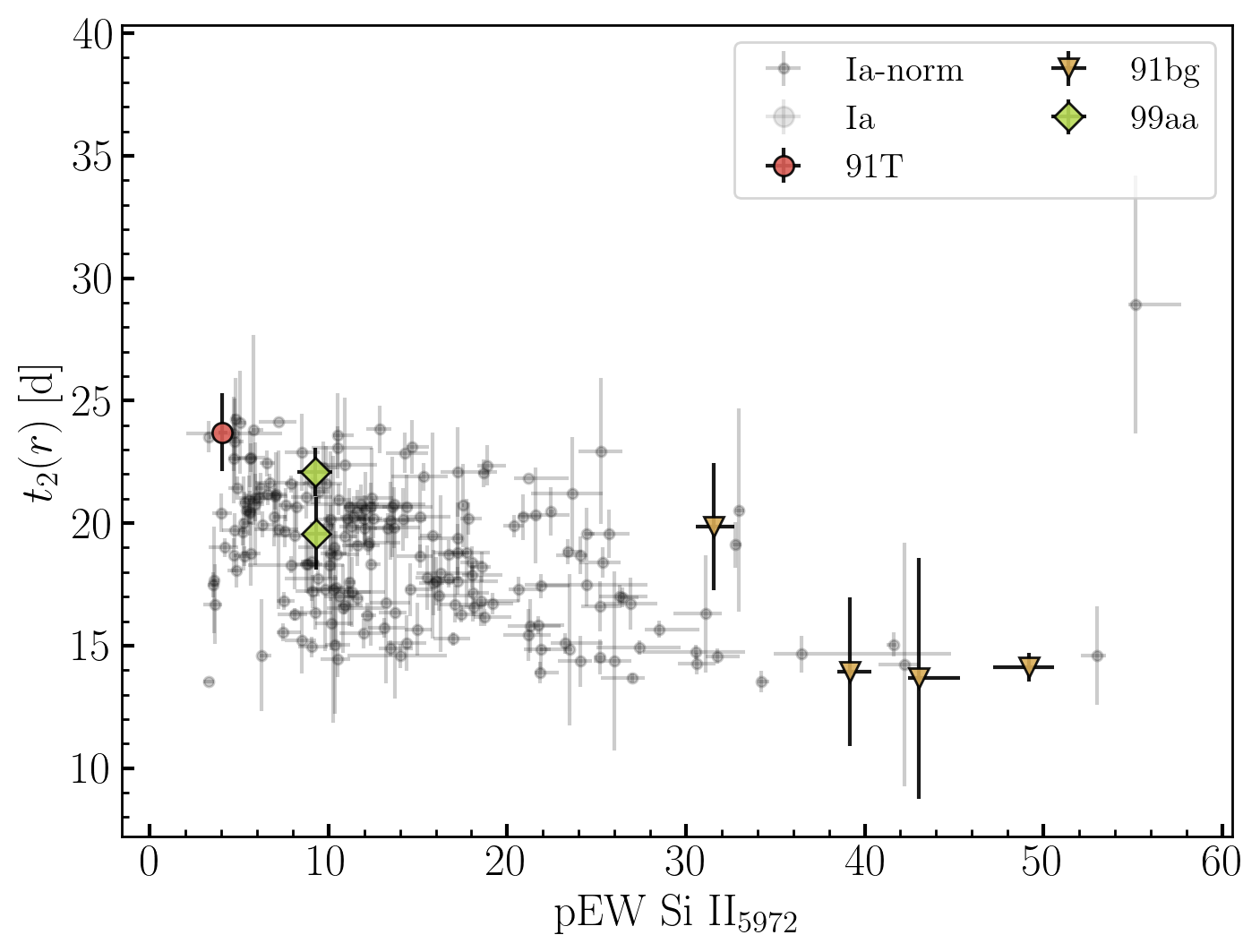}
    \caption{The time of $t_2(r)$ in the $r$ band as a function of the pEW of the 5972 \AA\ \SiII\ line.}
    \label{si_ratio_f2}
\end{figure}

\section{Conclusions}\label{conclusions}

We measured the time and strength of the secondary maximum in the \textit{r} band for a sample of 893 SNe~Ia using GP fits. Our main results are as follows:

\begin{enumerate}
    \item We find $>5\sigma$ correlations between $\Delta m_{15}(g)$ and the timing and strength of the secondary maximum, confirming that fainter and faster evolving SNe~Ia have earlier and weaker secondary maxima, and brighter, slower evolving SNe~Ia have later and stronger secondary maxima.
    \item We find that sub-classes of SNe Ia which are speculated to experience interaction with a CSM/envelope (03fg- and 02es-like, as well as Ia-CSM) have higher values of $\mathcal{F}_{r_2}$ than expected from the general trend of the sample, which can be explained by additional flux contribution from said interaction. Interestingly, we also find the 91T-like SNe Ia in general show higher $\mathcal{F}_{r_2}$-values than the trend, which could point to these SNe Ia experiencing some interaction. Several 02cx-like SNe Ia show higher $\mathcal{F}_{r_2}$-values which could be attributed to a bound remnant left behind after an incomplete explosion.
    \item We constrain the transparency timescales for our sample, and find that the distribution peaks at 35.9 $\pm$ 0.1 d, but there is a tail to lower transparency timescales. We fit the distribution with a bimodal Gaussian and find that the higher transparency timescale mode dominates, but there is a second component which peaks at 30.9 $\pm$ 0.4 d.
    \item The lower transparency timescale component of the bimodal Gaussian, peaking at 30.9 $\pm$ 0.4 d, is produced almost exclusively by SNe Ia residing in locally red environments. The main peak at 35.9 $\pm$ 0.1 d is produced by SNe Ia in both locally red and blue environments.
    \item The transparency timescales of most of the SNe Ia in our cosmological sample (97 per cent) are consistent with predictions from double detonation explosion models, whereas only 28 per cent are consistent with the predictions from delayed detonation models.
    \item We measure the total ejected mass using two methods: from the transparency timescale, which was estimated from $\mathcal{F}_{r_2}$, or from the correlation $M_{\textnormal{ej}}$ shows with $x_1$. Both methods are derived from bolometric light curves, and the distributions of $M_{\textnormal{ej}}$ we find from two methods are consistent, with medians of 1.3 $\pm$ 0.3 and 1.2 $\pm$ 0.2 \msun, respectively.
    \item We identified a spectral feature that sits in the $r$ band around 6500 \AA\ which begins to strengthen during the onset of the secondary maximum, and using a NLTE radiative transfer model we tentatively identify this feature as \FeII. 
    \item From the spectral time series of SN 1999by (a 91bg-like SN Ia) we find that these same features appear in the spectrum as early as 3 d post maximum, supporting the suggestion that the secondary maximum is not visible in 91bg-like SNe Ia because it coincides with the primary peak.
    \item We test the ability of the secondary maximum to work as a standardisation parameter and find that $\mathcal{F}_{r_2}$ shows a stronger correlation than $x_1$ with the colour-corrected Hubble residuals, with Pearson's correlation coefficients of $r\ =\ -$0.52 and $-$0.50, respectively. The rms dispersion the Hubble residuals dispersion is reduced by 0.004 mag when using $\mathcal{F}_{r_2}$ to correct compared to $x_1$.
    \item We find a slight non-linearity in the correlation between $x_1$ and $\mathcal{F}_{r_2}$ for the cosmological sub-sample. We fit the correlation with a broken line, and find a split at $x_1^0\ =\ -$0.5 $\pm$ 0.2. Interestingly, \cite{Senzel2024} notice a similar non-linearity between $x_1$ and the local SN colour, and \cite{Ginolin2024} find a split in $x_1$ at $-0.49\ \pm\ 0.06$ when plotted against the standardised Hubble residuals, which is consistent with our findings.
    \item We find that although the timing of the secondary maximum is thought to be primarily driven by temperature, the correlation between the pEW of \SiII\ line at 5972 \AA\ and the timing of the secondary maximum is weak ($r=-$0.38). 
    \item We find a $3.6\sigma$ correlation between the amount of radioactive nickel produced in the explosion and the timing of the secondary maximum. 
    \item We conclude the metallicity is a non-dominant parameter for variations of the secondary maximum because the $5\sigma$ trend between $t_2(r)$ and $\mathcal{F}_{r_2}$ opposes the trend expected from changes in metallicity predicted by \cite{Kasen2006}.
\end{enumerate}

The large sample of SNe Ia provided by ZTF has enabled us to study the secondary maximum in the $r$ band in great detail, both photometrically and spectroscopically. The 706 normal SNe Ia in our sample have allowed us to constrain the ranges of the parameters of the secondary maximum. We presented analysis for some of peculiar SN Ia sub-classes, but future studies with greater numbers of e.g. 91bg-like SNe Ia will be better able to constrain the range of when recombination occurs. The broken line fit between $x_1$ and $\mathcal{F}_{r_2}$, which resulted in two populations of SNe Ia split by $x_1^0\ =\ -0.5\ \pm\ 0.2$, should be investigated in greater detail in future studies as this may hold the key to determining if there are two main progenitor scenarios that result in slightly differing SN Ia properties. We re-iterate the importance of $\mathcal{F}_{r_2}$ because it can be measured for any SN Ia sub-class regardless of light curve morphology. Although in most cases it is easier to constrain $x_1$ than $\mathcal{F}_{r_2}$, it is interesting that $\mathcal{F}_{r_2}$ is better able to standardise SN Ia light curves. Standardisation using $\mathcal{F}_{r_2}$ may be useful for specific cases during LSST where a SN Ia has insufficient data on the rise to constrain $x_1$, but sufficient data later on to constrain the peak in the $r$ band and $\mathcal{F}_{r_2}$.

\section*{Acknowledgements}

MD, KM, GD, UB, and JHT are funded by the EU H2020 ERC grant no. 758638.
LJS acknowledge support by the European Research Council (ERC) under the European Union’s Horizon 2020 research and innovation program (ERC Advanced Grant KILONOVA No. 885281). 
LJS acknowledges support by Deutsche Forschungsgemeinschaft (DFG, German Research Foundation) - Project-ID 279384907 - SFB 1245 and MA 4248/3-1.
L.G. acknowledges financial support from the Spanish Ministerio de Ciencia e Innovaci\'on (MCIN) and the Agencia Estatal de Investigaci\'on (AEI) 10.13039/501100011033 under the PID2020-115253GA-I00 HOSTFLOWS project, from Centro Superior de Investigaciones Cient\'ificas (CSIC) under the PIE project 20215AT016 and the program Unidad de Excelencia Mar\'ia de Maeztu CEX2020-001058-M, and from the Departament de Recerca i Universitats de la Generalitat de Catalunya through the 2021-SGR-01270 grant.
This work has been supported by the research project grant “Understanding the Dynamic Universe” funded by the Knut and Alice Wallenberg Foundation under Dnr KAW 2018.0067.
This project has received funding from the European Research Council (ERC) under the European Union's Horizon 2020 research and innovation program (grant agreement n 759194 - USNAC).
AG acknowledges support from  {\em Vetenskapsr\aa det}, the Swedish Research Council, project 2020-03444.
SD acknowledges support from the Marie Curie Individual Fellowship under grant ID 890695 and a Junior Research Fellowship at Lucy Cavendish College. 
LH is funded by the Irish Research Council under grant number GOIPG/2020/1387.
T.E.M.B. acknowledges financial support from the Spanish Ministerio de Ciencia e Innovaci\'on (MCIN), the Agencia Estatal de Investigaci\'on (AEI) 10.13039/501100011033, and the European Union Next Generation EU/PRTR funds under the 2021 Juan de la Cierva program FJC2021-047124-I and the PID2020-115253GA-I00 HOSTFLOWS project, from Centro Superior de Investigaciones Cient\'ificas (CSIC) under the PIE project 20215AT016, and the program Unidad de Excelencia Mar\'ia de Maeztu CEX2020-001058-M.
Y.-L.K. has received funding from the Science and Technology Facilities Council [grant number ST/V000713/1].
Based on observations obtained with the Samuel Oschin Telescope 48-inch and the 60-inch Telescope at the Palomar Observatory as part of the Zwicky Transient Facility project. ZTF is supported by the National Science Foundation under Grants No. AST-1440341 and AST-2034437 and a collaboration including partners Caltech, IPAC, the Weizmann Institute of Science, the Oskar Klein Center at Stockholm University, the University of Maryland, Deutsches Elektronen-Synchrotron and Humboldt University, the TANGO Consortium of Taiwan, the University of Wisconsin at Milwaukee, Trinity College Dublin, Lawrence Livermore National Laboratories, IN2P3, University of Warwick, Ruhr University Bochum, Northwestern University and former partners the University of Washington, Los Alamos National Laboratories, and Lawrence Berkeley National Laboratories. Operations are conducted by COO, IPAC, and UW. SED Machine is based upon work supported by the National Science Foundation under Grant No. 1106171. The ZTF forced-photometry service was funded under the Heising-Simons Foundation grant \#12540303 (PI: Graham). 
This work made use of the Heidelberg Supernova Model Archive (HESMA), \url{https://hesma.h-its.org}.

\section*{Data Availability}

All the light curves and spectra used in this paper are publicly available (\url{https://github.com/ZwickyTransientFacility/ztfcosmoidr/tree/main/dr2}). Spectra taken from ZTF post-DR2 are available on WISEREP. The properties of the secondary maximum derived from our Gaussian Process fits are available in the online material.

\bibliographystyle{aa}
\bibliography{aanda} 

\end{document}